\documentclass{aa}  

\usepackage{graphicx}
\usepackage{txfonts}
\usepackage[colorlinks,allcolors=blue]{hyperref}

\usepackage{natbib}
\bibpunct{(}{)}{;}{a}{}{,} 
\usepackage{bm}
\usepackage{color}
\usepackage{ulem}
\usepackage{xspace}
\usepackage{float}
\usepackage{epstopdf}
\usepackage{CJKutf8}
\usepackage{hieroglf}
\usepackage[utf8]{inputenc}
\usepackage{placeins}

\ttfamily
\definecolor{mygray}{gray}{0.6}
\definecolor{magenta}{rgb}{0.858, 0.188, 0.478}
\definecolor{revise}{RGB}{0, 139, 0 }
\definecolor{revise2}{RGB}{255, 0, 0 }

\newcommand{\xxx}[1]{\textcolor{blue}{\textbf{xxx}\xspace}}

\newcommand{\fg}[1]{Fig.~\ref{fig:#1}}
\newcommand{\Fg}[1]{Figure~\ref{fig:#1}}

\newcommand{\eq}[1]{Eq.~(\ref{eq:#1})\xspace}
\newcommand{\beq}[1]{(Eq.~\ref{eq:#1})\xspace}

\newcommand{\tb}[1]{Table~\ref{tab:#1}\xspace}
\newcommand{\se}[1]{Sect.~\ref{sec:#1}\xspace}

\makeatletter
\def\uwave{\bgroup \markoverwith{\lower3.5\p@\hbox{\sixly \textcolor{red}{\char58}}}\ULon}
\font\sixly=lasy6 
\makeatother

\begin{document}

   \title{On the suppression of giant planet formation around low-mass stars in clustered environments}


   \author{Shuo Huang (\begin{CJK*}{UTF8}{gbsn}黄硕\end{CJK*}) {\bf\inst{1,2} }
          \and 
          Simon Portegies Zwart
          \inst{1} \and Maite J. C. Wilhelm \inst{1} 
          }
   \institute{Leiden Observatory, Leiden University, Einsteinweg 55, 2333 CC Leiden, The Netherlands \\
   \email{shuang@strw.leidenuniv.nl}
         \and
             Department of Astronomy, Tsinghua University, 100084 Beijing, China
        }
   \date{Received 26 June 2024 / Accepted 1 August 2024}

 
  \abstract
   {Current exoplanet formation studies tend to overlook the birth environment of stars in clustered environments. The effect of this environment on the planet-formation process, however, is important, especially in the earliest stage.}
   {We investigate the differences in planet populations forming in star-cluster environments through pebble accretion and compare these results with the planet formation around isolated stars. We try to provide potential signatures on the young planetary systems to guide future observation.
   }
   {We design and present a new planet population synthesis code for clustered environments. The planet formation model is based on pebble accretion and includes migration in the circumstellar disk. The disk's gas and dust are evolved in 1D simulations considering the effects of photo-evaporation of the nearby stars.
   }
   {Planetary systems in a clustered environment are different than those born in isolation; the environmental effects are important for a wide range of observable parameters and the eventual architecture of the planetary systems.  Planetary systems born in a clustered environment lack cold Jupiters compared to isolated planetary systems. This effect is more pronounced for low-mass stars ($\lesssim$0.2 $M_\odot$). On the other hand, planetary systems born in clusters show an excess of cold Neptune around these low-mass stars.}
   {In future observations, finding an excess of cold Neptunes and a lack of cold Jupiters could be used to constrain the birth environments of these planetary systems. Exploring the dependence of cold Jupiter's intrinsic occurrence rate on stellar mass provides insights into the birth environment of their proto-embryos.
   }


   \maketitle
%

\section{Introduction}
Since the first discovery of an exoplanet around a solar-type star \citep{MayorQueloz1995}, the rapidly growing number of detected exoplanets continues to refresh our understanding of planet formation. The abundance and diversity of known exoplanets provide the opportunity to examine planet formation theories on a statistical basis \citep{ZhuDong2021}. 

Planets are expected to form in the circumstellar disks surrounding young stars. The Atacama Large Millimeter/submillimeter Array (ALMA) has provided unprecedented views into these planet-forming disks \citep[e.g.][]{AndrewsEtal2018, LongEtal2018, CiezaEtal2019, OebergEtal2021}. ALMA continuum observations reveal the presence of (sub)millimeter-sized particles, commonly known as pebbles, within protoplanetary disks \citep[e.g.][]{AnsdellEtal2017}. These pebbles are the essential building blocks for planets and play a crucial role in the efficient accretion onto planet cores, facilitating their growth \citep[so-called pebble accretion,][]{OrmelKlahr2010, LambrechtsJohansen2012}. 

Pebble accretion is advantageous for the rapid growth of cold giants due to its short timescale \citep{JohansenEtal2019}. Planetesimal accretion can also grow planets, particularly in short Keplerian orbits \citep{JohansenBitsch2019}. As the core mass is massive enough, gas accretion commences and gas giants can be formed \citep[so-called core accretion][]{PerriCameron1974}. Another way to form a gas giant is direct Gravitational Instability \citep[GI,][]{Toomre1964, Boss1997}, but it requires a cold and massive disk that fragments and collapses. Once planets grow massive enough, they perturb the gas disk and exchange angular momentum, triggering planet migration \citep{LinPapaloizou1979, GoldreichTremaine1980}. These simplified planet formation theories allow for understanding the observed exoplanet demographics. With this objective, many planet population synthesis studies are performed \citep[e.g.][]{MordasiniEtal2009, IdaEtal2018, Chambers2018, ForganEtal2018, JohansenEtal2019, MiguelEtal2020, EmsenhuberEtal2021, JohnstonEtal2024, LauEtal2024i,ChakrabartyMulders2024}. 

The effects of the star cluster environment on planet formation are often not addressed in simplified planet formation models. The validity of such simplifications warrants examination from two perspectives: the importance and ubiquity of star clusters for the planet formation process.
Firstly, its importance is evident during planet formation. Protoplanetary disks, where planets' building blocks originate, can be influenced by the cluster environment. A compelling example is Proplyds, identified in the Orion Nebula Cluster (ONC) \citep{O'DellEtal1993, O'DellWen1994}. These objects are observed to lose mass due to intense external photo-evaporation (EPE), highlighting the impact of the cluster environment on planetary systems.
Various simulations are conducted addressing the EPE effect on protoplanet disks in the cluster. 
Secondly, the clustered nature of star formation is a common occurrence. Stars predominantly form within molecular clouds \citep{LadaLada2003}, which are extensive structures spanning tens to hundreds of light-years. The regions within these clouds undergo gravitational collapse that ultimately leads to the formation of star clusters \cite{2010ARA&A..48..431P}. 
Nevertheless, the relatively short disassociation timescale of clusters into the Galactic field \citep[$\sim100$ Myr,][]{2005A&A...441..117L,2010ARA&A..48..431P, KrumholzEtal2019} may result in a much less clustered distribution as we observed currently.

Star clusters exert various effects on protoplanetary disks, including external photoevaporation \citep[e.g.][]{WinterEtal2022,WilhelmEtal2023, QiaoEtal2023, WederEtal2023}, external heating \citep{NduguEtal2018}, and dynamical truncation \citep{PortegiesZwart2016i, NduguEtal2022}. Recent planet population synthesis studies have incorporated close truncation and external heating to investigate the behavior of planet demographics post-formation using planet population synthesis models \citep{NduguEtal2018, NduguEtal2022}. The impacts of external photoevaporation on the resulting planet population in clusters remain unclear.
In this paper, we employ our newly developed code, PACE using the
AMUSE framework \citep{PortegiesZwartEtal2009, PortegiesZwart2013456, PelupessyEtal2013}, to synthesize and simulate planet formation in the perturbed environments. 

The structure of the paper is outlined as follows: In \se{method}, we detail the processes involved in planet formation. \se{initial} describes our initial conditions and parameter selections. The results from planet population syntheses are analyzed in \se{default}. Additionally, we present our simplified analytical criteria for forming cold Jupiters and its validation against simulation results in \se{theory}. Finally, the discussion is presented in \se{discussion}, followed by the conclusion in \se{conclusion}.

\section{Methods}
\label{sec:method}
Our model includes four main physical processes: cluster evolution, proto-planet disk evolution, planet accretion, and planet migration as shown in \fg{pps_sketch}. Different sub-modules are bridged via VENICE (Wilhelm \& Portegies Zwart in prep.) using the AMUSE framework \citep{PortegiesZwartEtal2009, PortegiesZwart2013456, PelupessyEtal2013}. 

In the following subsections, we describe in detail how each sub-module works and interoperates, illustrating in \fg{pps_sketch}.
\begin{figure}
    \centering
    \includegraphics[width=0.5\textwidth]{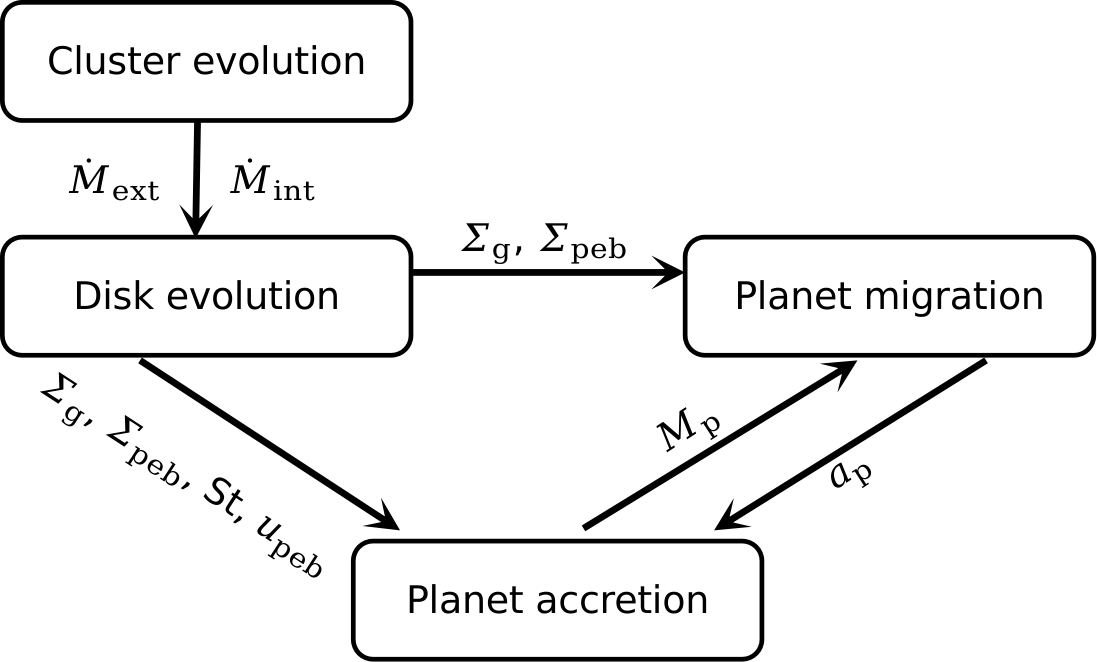}
    \caption{Sub-modules and most important exchanged quantities of our population synthesis model. The boxes indicate different physical modules. The arrows refer to the direction of data transported from one module to another. }
    \label{fig:pps_sketch}
\end{figure}

\subsection{Cluster evolution}
\label{sec:cluster_model}
The star formation in the molecular cloud is simulated. From the spatial position of stars, we can calculate and apply external photoevaporation to each planetary system. 
The EPE mass-loss rates are derived from interpolation on the FRIED grid \citep{HaworthEtal2018}. {Only stars more massive than 7 M$_\odot$ exert extreme ultraviolet (EUV, $E_\gamma = 13.6+\,\mathrm{eV}$) and far ultraviolet (FUV, $E_\gamma = 5.6 - 13.6\,\mathrm{eV}$) radiation to nearby stars.} Dynamical truncation is ignored as its effect is small compared to EPE \citep{WilhelmEtal2023}.

{We summarize the model used to evolve star formation and cluster evolution, following the approach outlined in {Wilhelm et al. in prep} \citep[employing methods similar to those described in][]{WilhelmEtal2023}. We use \texttt{TORCH} \citep{WallEtal2019, WallEtal2020} to simulate the collapse of the molecular cloud. The dynamical interactions between stars and the gas are considered using \texttt{BRIDGE} \citep{FujiiEtal2007, PortegiesZwartEtal2020} in the \texttt{AMUSE} framework \citep{PortegiesZwartEtal2009, PelupessyEtal2013}. } 

Initially, the star-forming cloud has a mass of $10^4$ M$_\odot$, a radius of 7 pc, and an ambient temperature of 30 K. The cloud features a Gaussian density profile and a turbulent velocity field with a Kolmogorov spectrum. Notably, our realization of the turbulent velocity field differs from that in \citet{WilhelmEtal2023} and has an initial virial ratio of 1. {The resulting stellar density in the cluster depends on the initial virial ratio.} The cloud is embedded in a uniform background medium with a density of 1.25 hydrogen nuclei cm$^{-3}$ and a temperature of 8000 K.   

In \fg{cluster_star} we present the cluster with its residual gas at an age of 2.43\, Myr after the start of the collapse of the giant molecular cloud. At this stage in the evolution, the cluster is composed of two separated structures, which are expected to merge eventually. The cluster has 6,890 stars with a total mass of 4019 M$_\odot$. 
The cluster is quite dense, with a 90\% Lagrangian radius of 1.69\,pc, a half-mass radius of 0.35\,pc, and a central density of $\sim 1.1\times 10^4$\, stars/pc$^3$.

We indicate the EPE of each disk in \fg{cluster_star} in colors. Circum-stellar disks close to the cluster density centers are more strongly affected by external photoevaporation than those further out, even though some of the most massive stars are orbiting the cluster periphery. Mass segregation has caused massive stars to sink toward the density center, and the core mass function is flat with an exponent of $\alpha\simeq 1.35$ compared to Salpeter ($\alpha = 2.35$). This flatter slope is consistent with the predicted core-mass function of a cluster in a state of core-collapse \cite{2011Sci...334.1380F}.
Circum-stellar disks that were prevented from escaping the cluster center are strongly affected by  FUV and EUV from the nearby massive stars. Strong radiation of nearby stars and the star formation process also drive a large cavity at the center of the molecular cloud. This reduces the effect of shielding the core population from FUV and EUV photons. 
The EPE of stars between $1.9$ and $7\, M_\odot$ are not calculated in the FRIED grid, and therefore not shown in \fg{cluster_star}.

\begin{table}
	\centering
	\caption{Default disk and planet parameters used during the planet formation. }
	\label{tab:parameters}
	\begin{tabular}{l|c|c} 
		\hline
		Description  & $\mathrm{Symbols}$    &  $\mathrm{Value(s)}$ \\
		\hline
        Metallicity     & $[\mathrm{Fe/H}]$  &  0.02        \\
        Disk-to-star mass ratio &  $M_\mathrm{d,0}/M_\star$   &  0.1  \\
        Mean molecular weight   & $\mu$  & 2.3          \\
        viscosity-$\alpha$ & $\alpha_\mathrm{t}$ & $10^{-4}$ \\
        accretion-$\alpha$ & $\alpha_\mathrm{acc}$ & $10^{-3}$ \\
        Adiabatic index & $\gamma$ & 7/5 \\
        Dust fragmentation velocity $[\mathrm{m/s}]$ & $v_\mathrm{frag}$ & 10 \\
        Minimum dust size $[\mathrm{cm}]$ & $a_0$ & 0.001 \\
        Envelope opacity $[\mathrm{m^2 kg^{-1}}]$ & $\kappa$ & 0.005 \\

		\hline
	\end{tabular}
\end{table}

The star-averaged EPE at each time and time-averaged EPE for each star are calculated in \fg{cluster_epe}, left and right panels respectively. The initial EPE is low because the most massive stars have not yet formed. 
Generally, the majority of disks have been externally evaporated effectively, with the mass loss rate of $\sim10^{-7}\, M_\odot\mathrm{yr^{-1}}$. Only 26 disks have an EPE smaller than $\sim10^{-9}\, M_\odot\mathrm{yr^{-1}}$: these stars are nearly isolated from the massive stars in the cluster and we expect {their circumstellar disks to disperse after $\gtrsim$ 10 Myr (so-called Peter Pan disk), which is much longer than typical disks \citep[][]{WilhelmPortegiesZwart2022}. }

\subsection{Disk model}
After a star forms, the conservation of angular momentum causes the nearby bound gaseous material to flatten and become a circumstellar disk. Stars with mass $<1.9\, M_\odot$ are assumed to form circumstellar disks in our model with $\sim10$ percent of the host mass. The further evolution of the disk is simulated following \cite{WilhelmEtal2023}. The model includes the viscous evolution of the gas disk and the dust radial drift. Our adopted disk-evolution model is 1-dimensional and does not account for asymmetric structures such as density spirals or warps. 

\subsubsection{Gas evolution}
The evolution of the gas profile in the disk bases on \texttt{VADER} \citep{KrumholzForbes2015}, which numerically integrates the $\alpha$-disk model \citep{ShakuraSunyaev1973, Lynden-BellPringle1974}. The disk accretion for $\alpha$-disk {$\lesssim10$ au is approximately}
\begin{equation}
\label{eq:mdot}
    \dot{M}_\mathrm{g} = 3\pi \nu \Sigma_\mathrm{g},
\end{equation}
where $v=\alpha_\mathrm{acc} h^2 \Omega$ is the viscosity, $\Omega$ is the Keplerian orbital frequency, and $h=\sqrt{k_\mathrm{B}T/\mu m_H}$ is the disk scale height. As pointed out by \cite{JohansenEtal2019}, the $\alpha$-disk model would not be valid if the disk accretion is driven by disk winds \citep{BaiStone2013}. We therefore simply parameterize $\alpha_\mathrm{acc}$ as a dimensionless representation of the disk accretion flux, instead of the turbulent viscosity parameter that is typically used.

External photoevaporation (EPE) causes the reduction of the disk mass from the outer disk edge inward. The amount of mass removed depends on the irradiation due to the other stars in the cluster (\se{cluster_model}). {The EPE mass loss rates are updated every 10 kyr. Given that the evolution of the cluster structure occurs on much longer timescales and is not affected by the planet formation processes, this approach does not compromise the validity of our conclusions.}

The effect of internal photo-evaporation (IPE) depends on the X-ray luminosity from the central star. The X-ray luminosity-mass relation of stars is fitted by \cite{FlaccomioEtal2012} using the {Chandra} observation on stars in the Orion Nebula Cluster:
\begin{equation}
    L_\mathrm{X} = 10^{1.7\log_{10}(M_\star/M_\odot) + 30} \,\mathrm{erg/s}
\end{equation} 
The IPE rate is calculated through:
\begin{equation}
    \log_{10}\left(\frac{\dot{M}_\mathrm{IPE}}{1\,M_\odot\mathrm{yr^{-1}}}\right) = A_\mathrm{L}\exp\left[\frac{(\ln(\log_{10}(\frac{L_X}{1\,\mathrm{erg/s}}))-B_\mathrm{L})^2}{C_\mathrm{L}}\right]+D_\mathrm{L},
\end{equation}
where $A_\mathrm{L}=-2.7326$, $B_\mathrm{L}=3.3307$, $C_\mathrm{L}=-2.9868$ and $D_\mathrm{L}=-7.2580$ \citep{PicognaEtal2019}. 

\subsubsection{Dust evolution}
\label{sec:dust}

\begin{figure}
    \centering
    \includegraphics[width=\columnwidth]{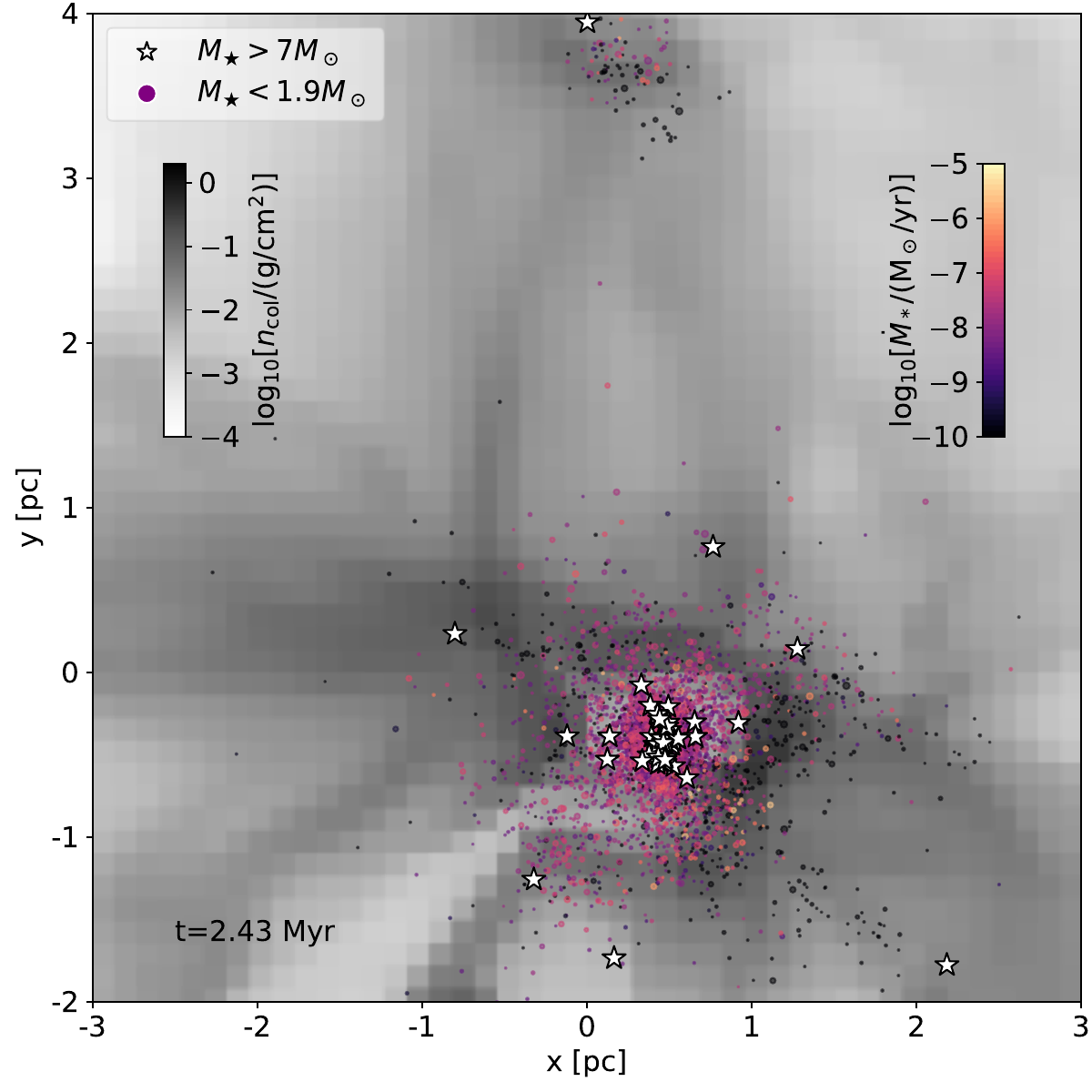}
    \caption{Spatial distribution of stars born in the collapsing molecular cloud in the final snapshot. The star symbol indicates the position of massive background stars ($>7M_\odot$) that emit FUV and EUV radiation. Dots represent the relative low-mass stars ($<1.9 M_\odot$) that host disks. The size of the dot refers to the mass of the star, smaller dots are for smaller stars. The color of the dot indicates the external radiation strength and we convert it to the EPE of the disk, the magma color map is for the high EPE and the black is for the low EPE. The background gray color map of the grids represents the column density of the molecular cloud. }
    \label{fig:cluster_star}
\end{figure}

We adopt the two-population model based on \cite{BirnstielEtal2012}. It utilizes two characteristic dust populations with small-size and large-size grains of radius $a_0$ and $a_1(t)$, respectively. Initially, all dust consists of small grains with $a_1(t=0)=a_0$. Over time, the small grain size $a_0$ remains fixed, while $a_1$ increases exponentially on the timescale of
\begin{equation}
\tau_\mathrm{grow} = \xi (Z_0\Omega_\mathrm{K})^{-1}.
\end{equation}
Here, $Z_0$ is the dust-to-gas ratio at orbital frequency $\Omega_\mathrm{K}$, with $\xi=10$ \citep{KrijtEtal2016}. The final grain size $a_1$ is limited by radial drift, fragmentation, and drift-induced fragmentation. The mass fraction of the large dust population is
\begin{equation}
\begin{aligned}
f_{\mathrm{m}}
     & = \left\{\begin{array}{ll}
        0.97 & \mathrm{if}\, \mathrm{radial \,drift \,limits \,dust \,size}, \\
        0.75 
         & \mathrm{otherwise}.
    \end{array}
    \right.
\end{aligned}
\end{equation}

{The two-population model allows for shorter computing times, which is advantageous in planet population synthesis studies. However, a more precise dust size evolution model, such as \texttt{Dustpy} \citep{StammlerBirnstiel2022}, costs more computational time.}


\begin{figure*}
    \centering
    \includegraphics[width=1.1428\columnwidth]{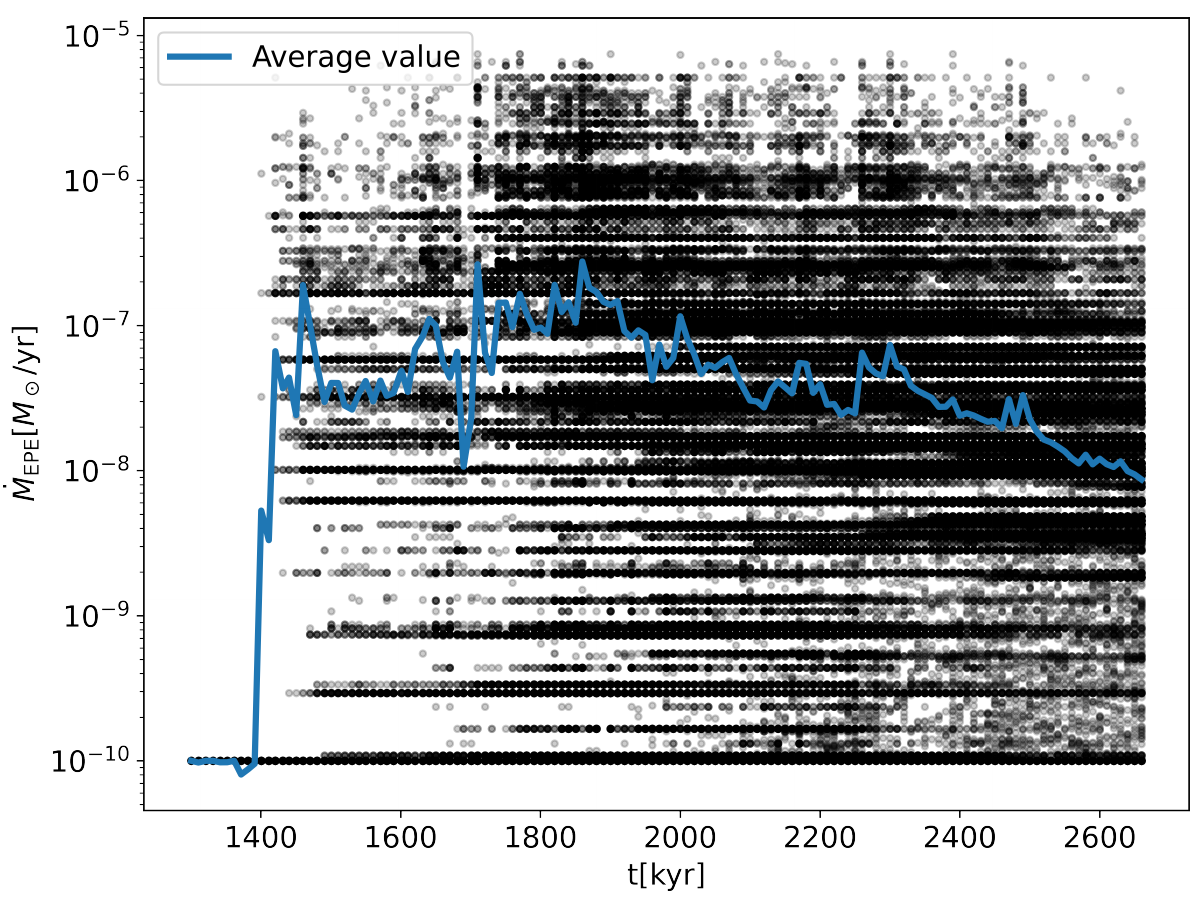}
    \includegraphics[width=0.8571\columnwidth]{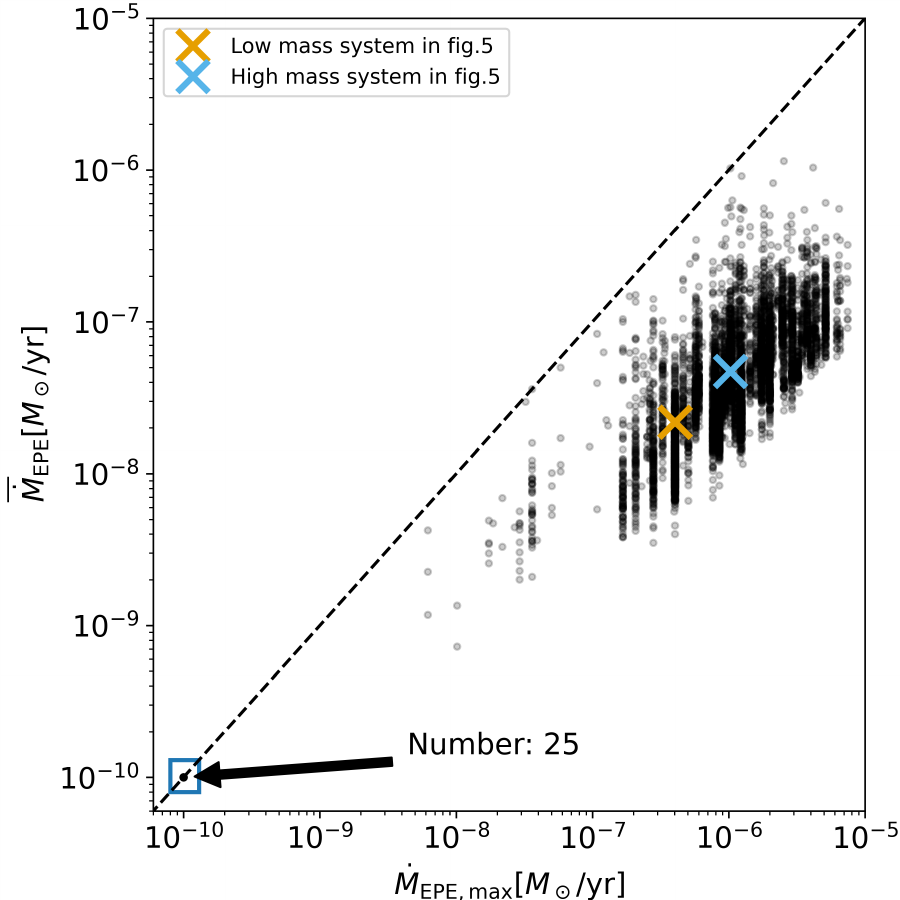}
    \caption{External photo-evaporation rate of disks in the cluster. The left panel shows the external photo-evaporation rate of each disk (black dots) and the averaged values over all disks (blue line) at different simulation times. The right panel shows the maximum external photo-evaporation rate versus the averaged value over the cluster lifetime of each disk. We also indicate the location and the number of isolated stars with their external photo-evaporation rates remaining at the lowest limit. In addition, the yellow and blue crosses are low and high mass systems, which are discussed further in \se{two-example}. }
    \label{fig:cluster_epe}
\end{figure*}

We incorporate the effect of EPE on the dust as well.
Gas evaporates from the disk's midplane due to EPE. The gas flow carries small grains, while larger grains remain unaffected. The maximum grain size ($a_\mathrm{entr}$) that can be entrained with the EPE gas flow ranges from 0.1 $\mu \mathrm{m}$ to 1 $\mathrm{mm}$ \citep[see][for detailed calculations]{FacchiniEtal2016}. {In this model, we assume a size distribution of the dust with power law $p=3.5$.} The fraction of dust smaller than $a_\mathrm{entr}$ is calculated and subtracted \citep{SellekEtal2020b}. 

\subsection{Planet growth}

Planet growth in this paper follows core accretion \citep{PerriCameron1974}. The planet first forms the core, and it starts to accrete gas when it is massive enough. 
\subsubsection{Pebble accretion}
Pebble accretion \citep{OrmelKlahr2010, LambrechtsJohansen2012} is efficient in growing planets, especially for Ceres- or Moon-sized planet embryos \citep{VisserOrmel2016, JohansenLambrechts2017}. The embryo grows by accreting pebbles at a rate:
\begin{equation}
    \dot{M}_\mathrm{peb} = \varepsilon 2\pi r u_\mathrm{peb} \Sigma_\mathrm{peb}.
\end{equation}
Here $\varepsilon$ is the pebble accretion efficiency. We calculated the efficiency following \cite{OrmelLiu2018}\footnote{The python script for calculating $\varepsilon$ is available at: \url{https://github.com/chrisormel/astroscripts}}. The pebble-drift velocity is $u_\mathrm{peb}$. Both $\varepsilon$ and $u_\mathrm{peb}$ depend on pebble Stokes number $\mathrm{St}$. For self-consistency, we calculate the mass-weighted Stokes number
\begin{equation}
    \mathrm{St} = (1-f_\mathrm{m})\mathrm{St}_0+f_\mathrm{m} \mathrm{St}_1,
\end{equation}
and the mass-weighted drift velocity $u_\mathrm{peb}$
\begin{equation}
    u_\mathrm{peb} = (1-f_\mathrm{m})v_0+f_\mathrm{m} v_1,
\end{equation}
from the two-population dust model (\se{dust}) for the pebble flux. 
The stokes number and drift velocity for the small dust population are represented by $\mathrm{St}_0$ and $v_0$, respectively, while $\mathrm{St}_1$ and $v_1$ for large dust grains. 


Once a planet grows sufficiently large to induce a pressure bump exterior to its orbit, it halts the incoming pebble flux and, as a consequence, pebble accretion terminates. We adopt the 3D pebble isolation mass from \citep{BitschEtal2018}
\begin{equation}
\begin{aligned}
\label{eq:pebbleiso}
    M_\mathrm{iso} = & 25 M_\oplus \left(\frac{h}{0.05}\right)^3\left[0.34\left(\frac{-3}{\log_{10}\alpha_t}\right)^4+0.66\right] \\
    & \times \left[1- \frac{\partial \ln P/\partial \ln r+2.5}{6}\right]\frac{M_\star}{M_\odot}.
\end{aligned}
\end{equation}
Here $\partial P/\partial r$ is the local pressure gradient of the radial gas disk, and $\alpha_\mathrm{t}$ is the viscous alpha-parameter.


\begin{figure*}
    \centering
    \includegraphics[width=\columnwidth]{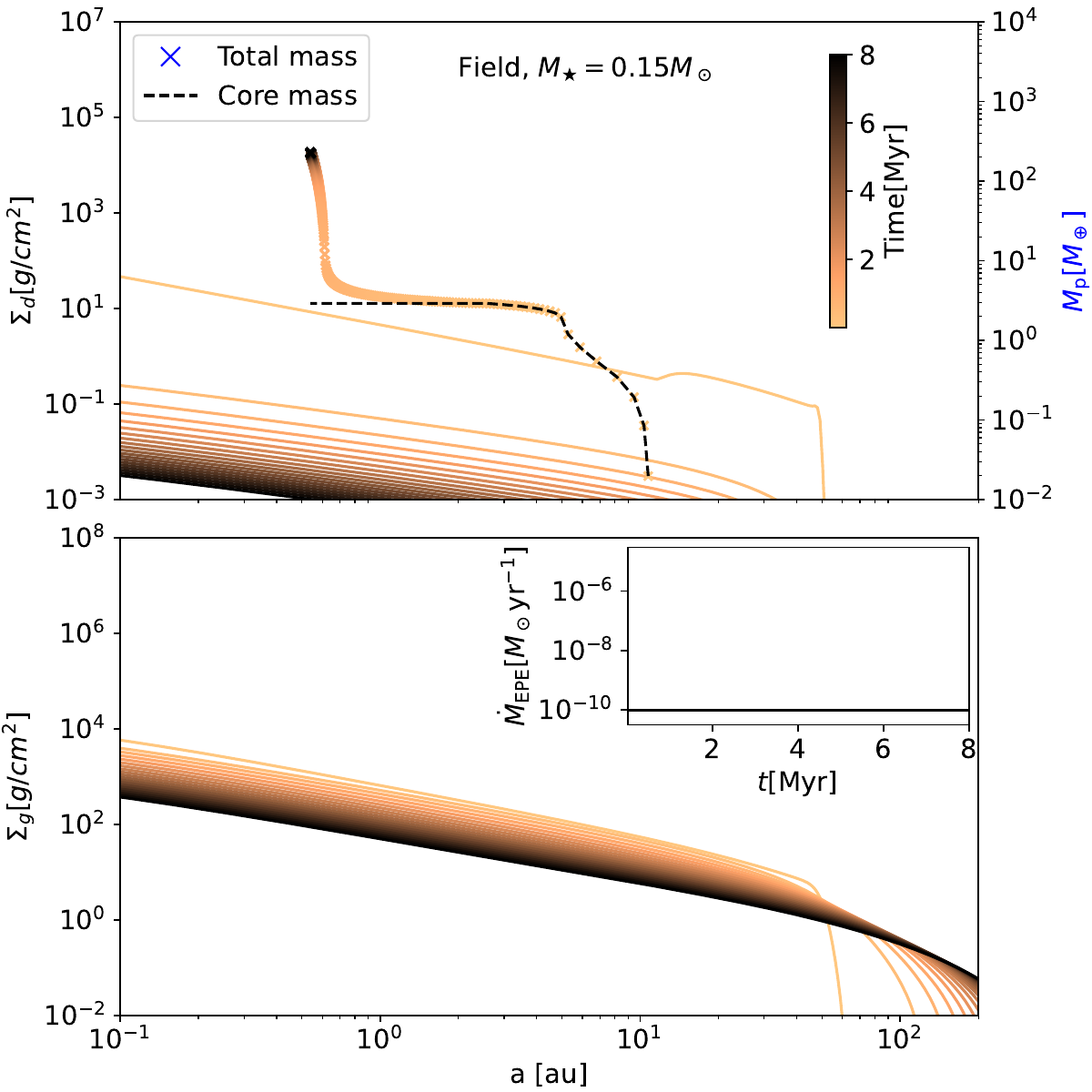}
    \includegraphics[width=\columnwidth]{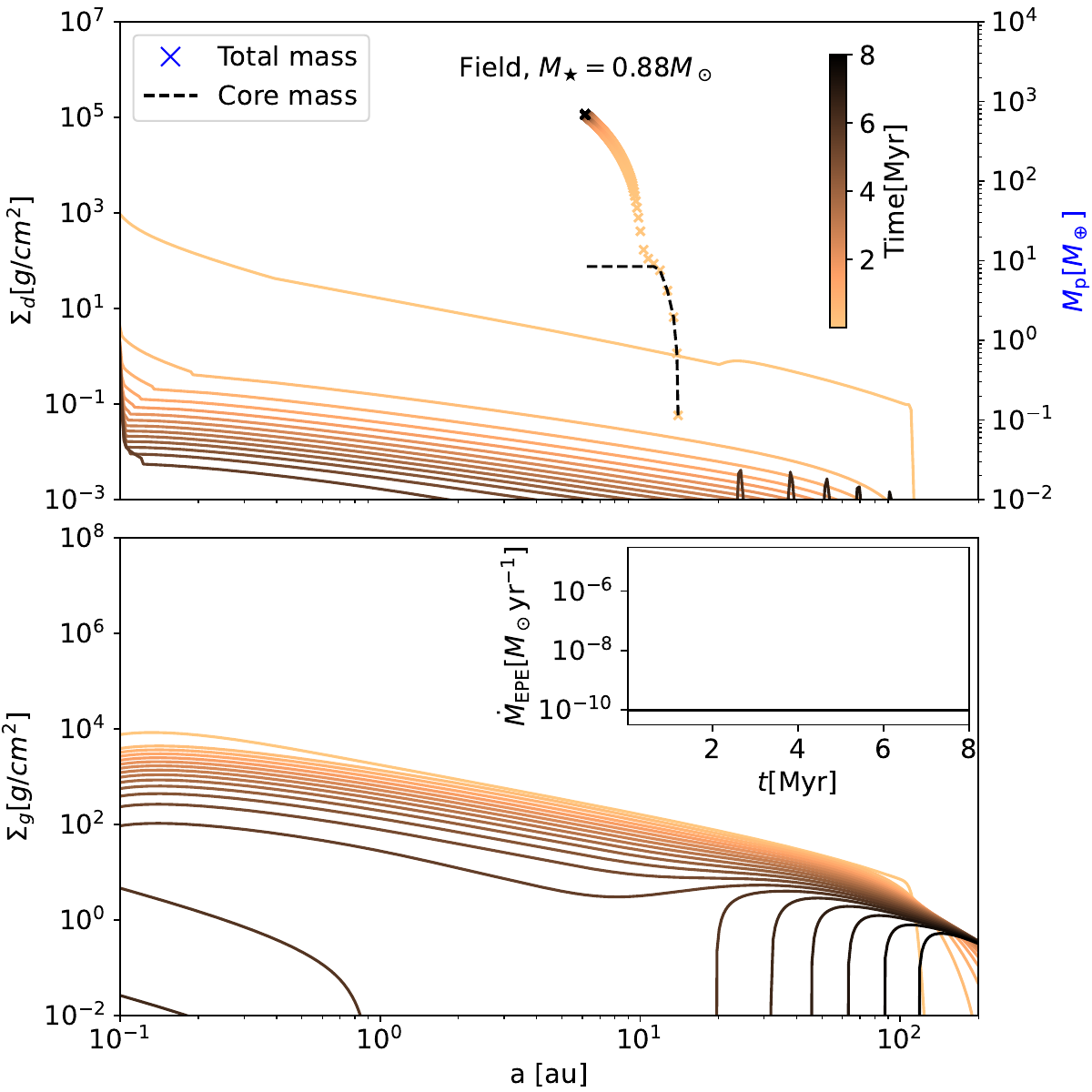}
    \caption{Comparison of disk evolution and planet growth tracks between a low-mass star ($0.15M_\odot$, left panels) and a relatively high-mass star ($0.88M_\odot$, right panels) in an isolated environment ($\dot{M}_\mathrm{EPE}=10^{-10} M_\odot\mathrm{yr^{-1}}$).In the top panels, we display the evolution of dust surface density (left axis) alongside planet mass (right axis). The planet's total mass and core mass are tracked separately using cross dots and dashed lines, respectively. The bottom panels show the evolution of gas surface density. The inset in the top right corner illustrates the history of external photo-evaporation rates experienced by the system. Both the top and bottom panels in each column share the same color bar, indicating the age of the system. }
    \label{fig:field_vs_cluster_example}
\end{figure*}

\subsubsection{Gas accretion}
As pebble accretion stops when a planetary core reaches the pebble isolation mass, hydrodynamic pressure no longer supports the gas envelope around the core against planetary gravity \citep{LambrechtsEtal2014}. If the supply of gas is sufficient, the gas accretion rate onto the gas giant is regulated by Kelvin–Helmholtz quasi-static contraction of the gas envelope \citep{IkomaEtal2000}: 
\begin{equation}
\label{eq:KH}
    \dot{M}_\mathrm{KH}=10^{-5}\left(\frac{M_\mathrm{p}}{M_\oplus}\right)^4\left(\frac{\kappa}{0.1\,\mathrm{m^2kg^{-1}}}\right)^{-1}M_\oplus\mathrm{yr^{-1}}.
\end{equation}
Here $\kappa$ is the opacity of the planet envelope. However, the gas accretion rate can later be limited by the amount of gas flowing into the Hill sphere at a rate of \citep{TanigawaTanaka2016}: 
\begin{equation}
\label{eq:hillacc}
    \dot{M}_\mathrm{Hill}=\frac{0.29}{3\pi h^{4}}\left(\frac{M_\mathrm{p}}{M_\star}\right)^{4/3}\frac{\dot{M}_\mathrm{g}}{\alpha_\mathrm{acc}}\frac{\Sigma_\mathrm{gap}}{\Sigma_\mathrm{g}},
\end{equation}
where $\Sigma_\mathrm{gap}/\Sigma_\mathrm{g}\cong(1+0.04K)^{-1}$ is the fitted depth of the gap that a massive planet creates in the gas disk. This estimate matches the results of hydrodynamic simulations from previous studies \citep[e.g.,][]{DuffellMacFadyen2013, KanagawaEtal2015, FungChiang2016}.
The gap depth factor $K$ is
\begin{equation}
    K=\left(\frac{M_\mathrm{p}}{M_\star}\right)^2 h^{-5}\alpha_\mathrm{t}^{-1}.
\end{equation}
One other limitation of the gas flow is the global disk-accretion rate $\dot{M}_\mathrm{g}$. We therefore set the gas accretion rate onto giant planets as the minimum value of all the above:
\begin{equation}
    \dot{M}_\mathrm{GA} = \min \left[\dot{M}_\mathrm{KH}, \dot{M}_\mathrm{Hill}, \dot{M}_\mathrm{g}\right].
\end{equation}
Such gas accretion methodology is the same as the one implemented in \cite{IdaEtal2018} and \cite{JohansenEtal2019}. 

{The disk accretion rate $\dot{M}_\mathrm{g}$ used in this study is an approximation (\eq{mdot}), chosen for the simplicity of calculations. The exact disk accretion rate depends on the detailed processes of disk evaporation (e.g., IPE and EPE), and may differ from the rate we used.}

\subsection{Planet migration}
When planets grow to Moon-mass bodies, they begin to gravitationally interact with the gas disk, exchanging angular momentum and thereby experiencing natural migration within the disk. We adopt the locally isothermal Type~I migration prescription as outlined in \cite{PaardekooperEtal2010, PaardekooperEtal2011}. This migration model considers the total torque, which is the combined effect of Lindblad torque ($\Gamma_\mathrm{L}$) and corotation torque ($\Gamma_\mathrm{C}$), both of which scale with the gas surface density. We simply refer to the appendix of \cite{IzidoroEtal2021} for a detailed summary of the migration prescription.

As planets grow in mass, their migration transitions into the Type II regime. The planets begin to open gaps in the gas disk. \cite{KanagawaEtal2018} discovered that the torque exerted on these gap-opening planets is suppressed, dependent on the surface density at the gap's bottom. Consequently, the migration timescale is generalized to
\begin{equation}
\tau_a = -\frac{1+0.04K}{\gamma_\mathrm{L}+\gamma_\mathrm{C}\exp{(-K/K_\mathrm{t})}}\tau_0(r_\mathrm{p}).
\label{eq:migration}
\end{equation}
Here $\gamma_\mathrm{L}$ and $\gamma_\mathrm{C}$ are the strengths of the Lindblad torque ($\gamma_\mathrm{L}=\Gamma_\mathrm{L}/\Gamma_0$), and corotation torque ($\gamma_\mathrm{C}=\Gamma_\mathrm{C}/\Gamma_0$) relative to the nominal torque $\Gamma_0=(q/h)^2\Sigma_\mathrm{g} r_\mathrm{p}^4\Omega_\mathrm{p}^2$. The nominal migration timescale is given by $\tau_0=r_\mathrm{p}^2\Omega_\mathrm{p}M_\mathrm{p}/2\Gamma_0$. For low viscosity conditions ($\alpha_\mathrm{t}<10^{-2}$), the corotation torque cutoff parameter $K_\mathrm{t}$ is fitted to a value of 20. We utilize \eq{migration} to compute the migration timescale for planets. 
When a planet is of low mass ($K\sim 0$), the migration timescale follows the Type I prescription: $\tau_\mathrm{a,I}=(\gamma_\mathrm{L}+\gamma_\mathrm{C})^{-1}\tau_0$.



\begin{figure*}
    \centering
    
    \includegraphics[width=\columnwidth]{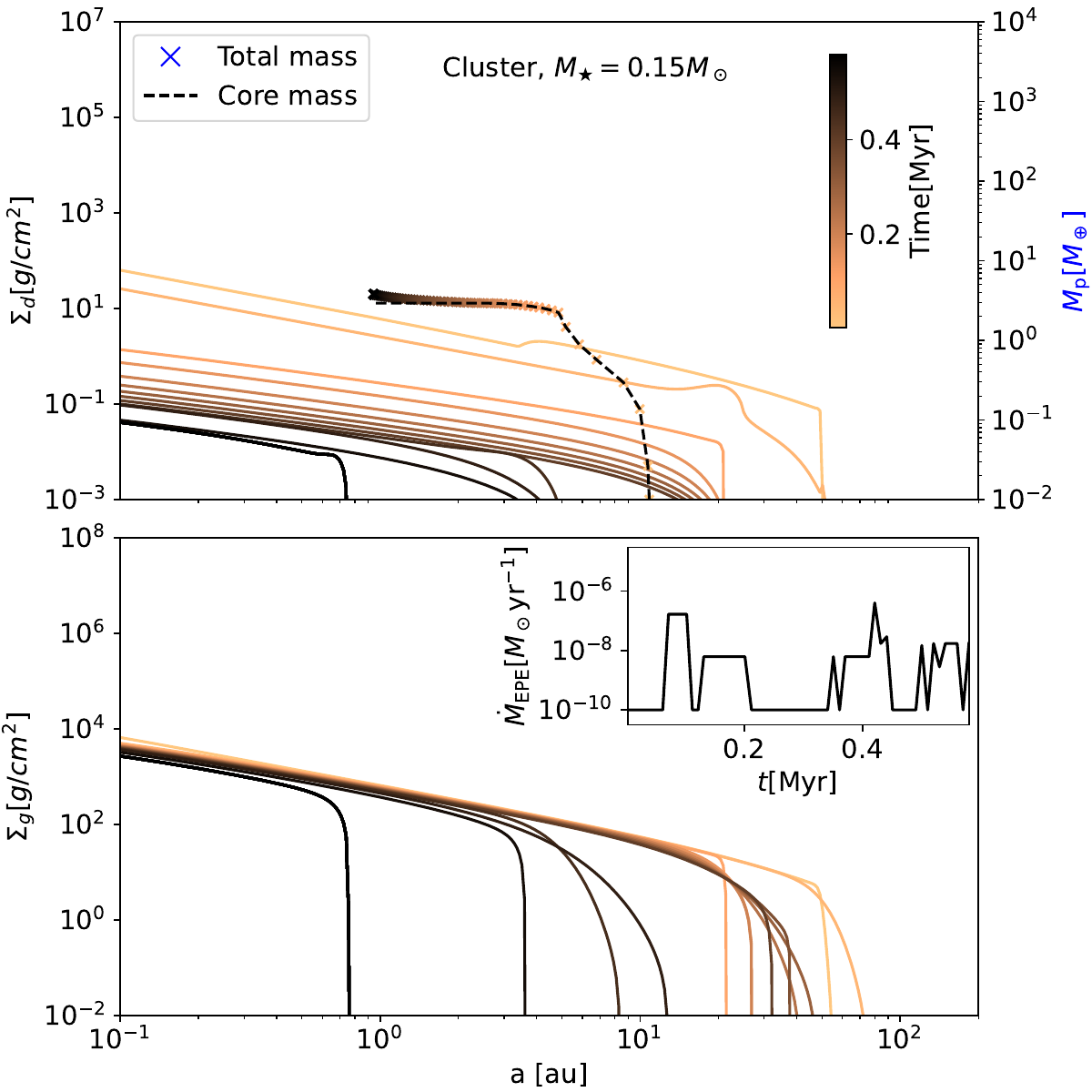}
    \includegraphics[width=\columnwidth]{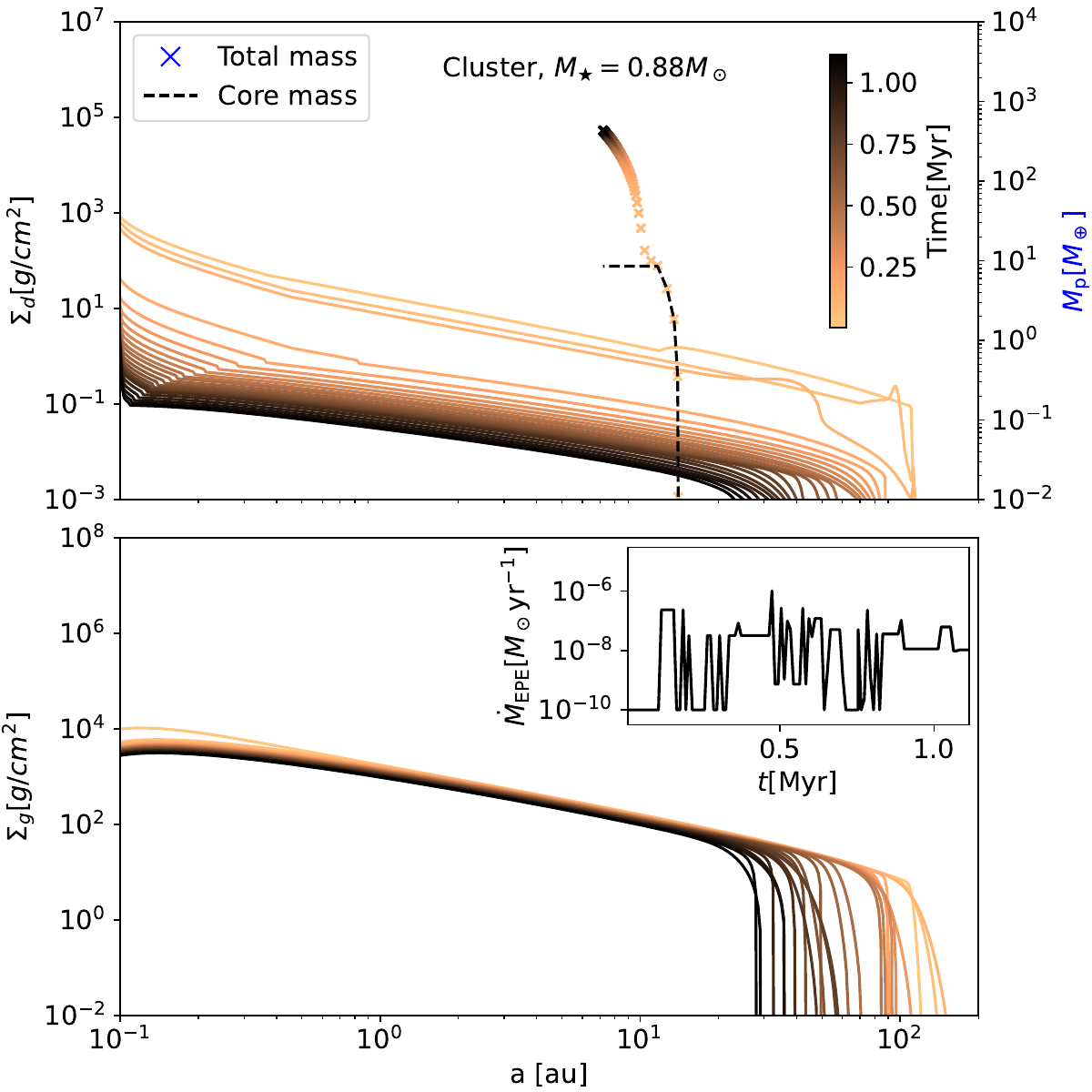}
    \caption{Same as \fg{field_vs_cluster_example}, but in the cluster. The star masses are $0.15M_\odot$ (left) and $0.88M_\odot$ (right). The insets on the top right of the bottom panels indicate the external photo-evaporation rate history the system experienced in the cluster. }
    \label{fig:field_vs_cluster_example_massive}
\end{figure*}

\section{Initial conditions and parameter choices}
\label{sec:initial}
The parameters utilized in our model are outlined in \tb{parameters}, with brief explanations provided below.

For the gas disk, we adopt initial conditions and boundary conditions similar to \cite{WilhelmPortegiesZwart2022}. The gas disk's initial density profile follows a self-similar formulation \citep{Lynden-BellPringle1974}:
\begin{equation}
\Sigma_\mathrm{g}(t=0) = \frac{M_\mathrm{d}}{2\pi R_\mathrm{out}(1-e^{-1})}\frac{\exp(-r/R_\mathrm{out})}{r},
\end{equation}
where $M_\mathrm{d}=0.1M_\star$ is the initial disk mass. However, our model diverges from self-similarity due to the inclusion of external photoevaporation (EPE) and internal photoevaporation (IPE). The outer disk radius $R_\mathrm{out}$ is determined by:
\begin{equation}
\label{eq:rdisk}
R_\mathrm{out} = 117\left(\frac{M_\star}{M_\odot} \right)^{0.45} \mathrm{au},
\end{equation}
{which is used in \citep{WilhelmEtal2023}}. 

The inner boundary of the disk is set by the inner disk radius, which is randomly distributed between $2$ and $12$ days following a normal distribution: $\log_{10}(P_\mathrm{in}/1,\mathrm{day})\sim\mathcal{N}(\mu_\mathrm{in},\sigma_\mathrm{in}^2)$, with $\mu_\mathrm{in}=\log_{10}2$ and $\sigma_\mathrm{in}=0.5$ \citep{BatyginEtal2023}.

The disk temperature profile is assumed to be
\begin{equation}
    T = 100 \left(\frac{M_\star}{M_\odot}\right)^{1/4}\left(\frac{r}{1\,\mathrm{au}}\right)^{-1/2} \,\mathrm{K},
\end{equation}
which has a similar radial power law as the radiation-dominated temperature case \citep[the power-law is $-3/7$ in][]{ChiangGoldreich1997}. Such low temperature is consistent with the cases in magnetically accreting laminar disks \citep{MoriEtal2021}. 

In our model, we set $\alpha_\mathrm{acc}=10^{-3}$ and $\alpha_\mathrm{t}=10^{-4}$, with $\alpha_\mathrm{acc}$ representing gas angular momentum via a viscous gas evolution model and $\alpha_\mathrm{t}$ denoting viscous-$\alpha$. The value of $\alpha_\mathrm{acc}$ is motivated by relatively high observed disk accretion rates \citep[e.g.,][]{Rafikov2017}. Lower value of $\alpha_\mathrm{t}$ comes from low disk viscosity in some observed disks \citep[e.g.,][]{FedeleEtal2018, FlahertyEtal2020, Portilla-ReveloEtal2023}. The adiabatic index of the disk is set to 7/5.

The initial and minimum dust grain size $a_1(t=0)$ and $a_0$ are both 0.001 cm. The impact on the final results is negligible due to rapid dust growth timescales compared to orbital timescales. The dust fragmentation velocity is $v_\mathrm{frag}=10\,\mathrm{m/s}$.

One planet seed per disk is initialized outside the iceline, with the semimajor axis following a uniform distribution $\log_{10}(a_\mathrm{p,0}/1\,\mathrm{au})\sim\mathcal{U}(\log_{10}(R_\mathrm{ice}/1\,\mathrm{au}), (R_\mathrm{out}/1\,\mathrm{au}))$. The iceline is located at $R_\mathrm{ice} \approx 0.67 \left({M_\star}/{M_\odot} \right)^{0.5}\,\mathrm{au}$, where the disk temperature is 150 K.
Each planet seed is of 0.01 $M_\oplus$, optimizing pebble-accretion efficiency over planetesimal accretion \citep{VisserOrmel2016, JohansenLambrechts2017}. 


\section{Results}
\label{sec:default}
\subsection{Planet formation for two example stars}
\label{sec:two-example}

We first show the planet growth track and the disk evolution around a 0.15 $M_\odot$ star and a 0.88 $M_\odot$ star in isolation in \fg{field_vs_cluster_example}, with $\dot{M}_\mathrm{EPE}$ fixed to $10^{-10}M_\odot\mathrm{yr^{-1}}$ for both calculations. Both systems are simulated for 8 Myr. 

The evolution of gas and dust follows distinct pathways around low-mass and high-mass stars but also shows some common features. In the early stages, gas profiles evolve viscously, leading to an extended disk outer edge. By around 4 Myr, the high-mass star, with its stronger internal photoevaporation (IPE) rate, drives a gas gap out to about 10 au. This gap's size is determined by the balance of radiation intensity, gas density, and temperature profile. Such a gap does not appear in the disk of low-mass stars until later in the simulation. {The stellar accretion rates are compared with the observed disks in \citep{WilhelmPortegiesZwart2022}.} In the initial snapshots of both systems, dips in dust surface density are observed at 10 and 20 au. These occur because small dust particles beyond these locations have insufficient time to grow larger due to longer growth timescales at greater distances from the star. The dust disk around the high-mass star shows peaks at about 50 au, attributed to a pressure maximum in the gas profile induced by IPE. Over time, the dust profile diminishes due to rapid radial drift. 

\begin{figure*}
    \centering
    \includegraphics[width=\columnwidth]{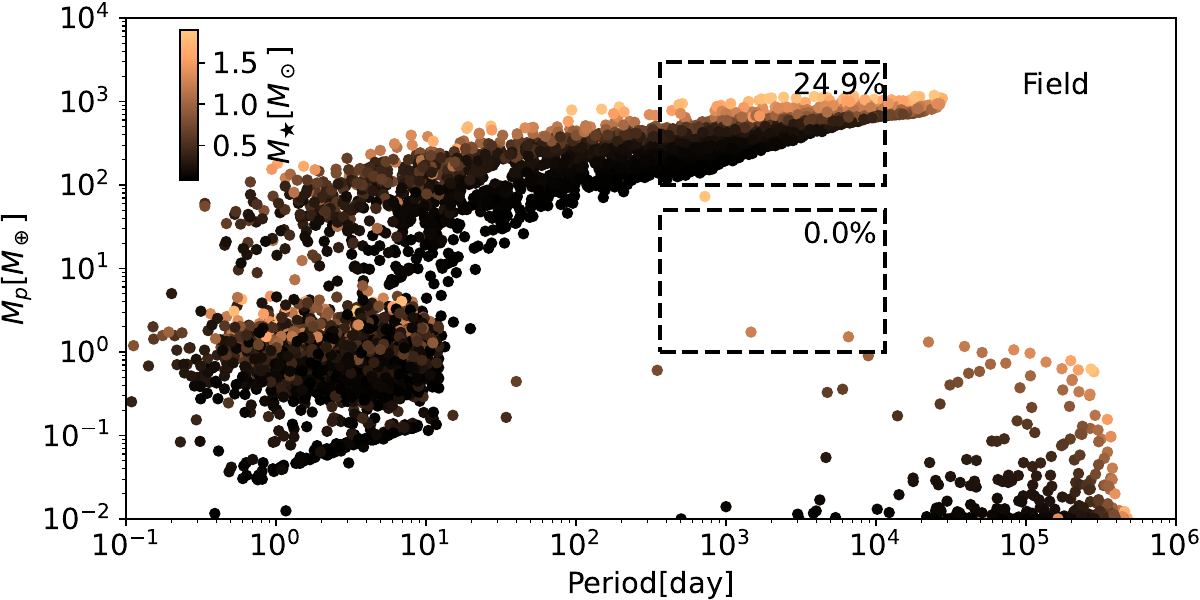}
    \includegraphics[width=\columnwidth]{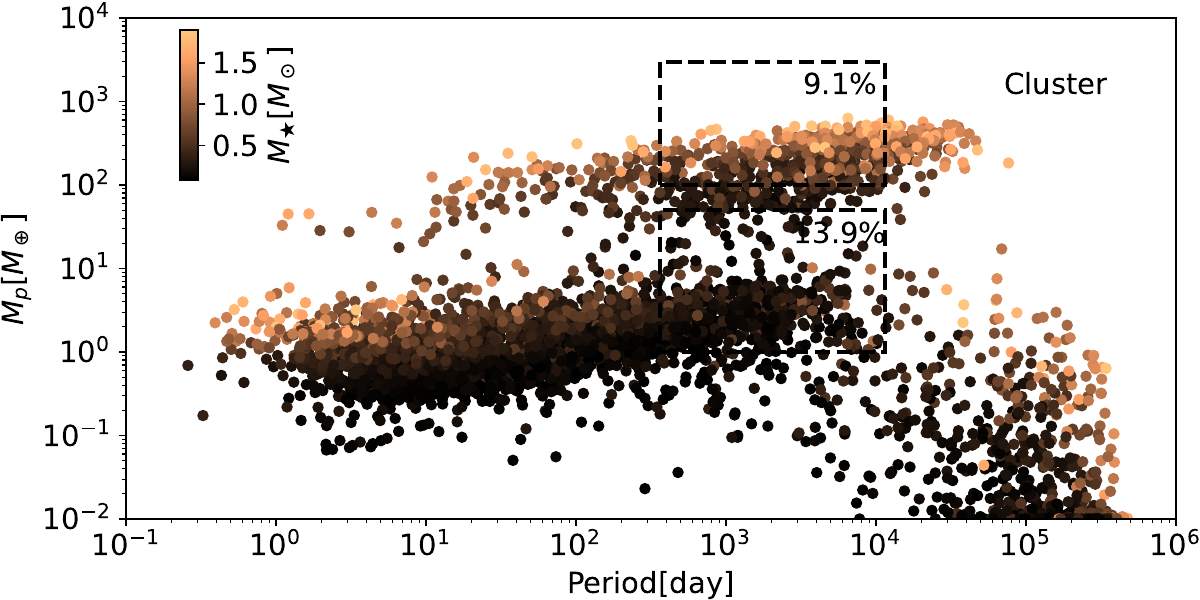}
    \includegraphics[width=\columnwidth]{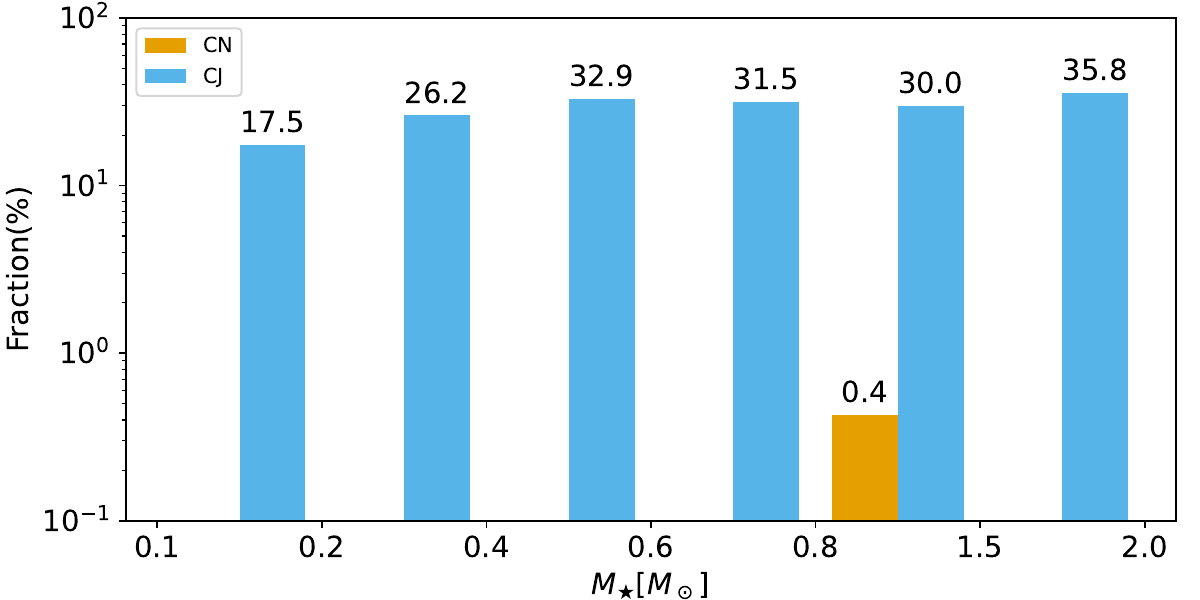}
    \includegraphics[width=\columnwidth]{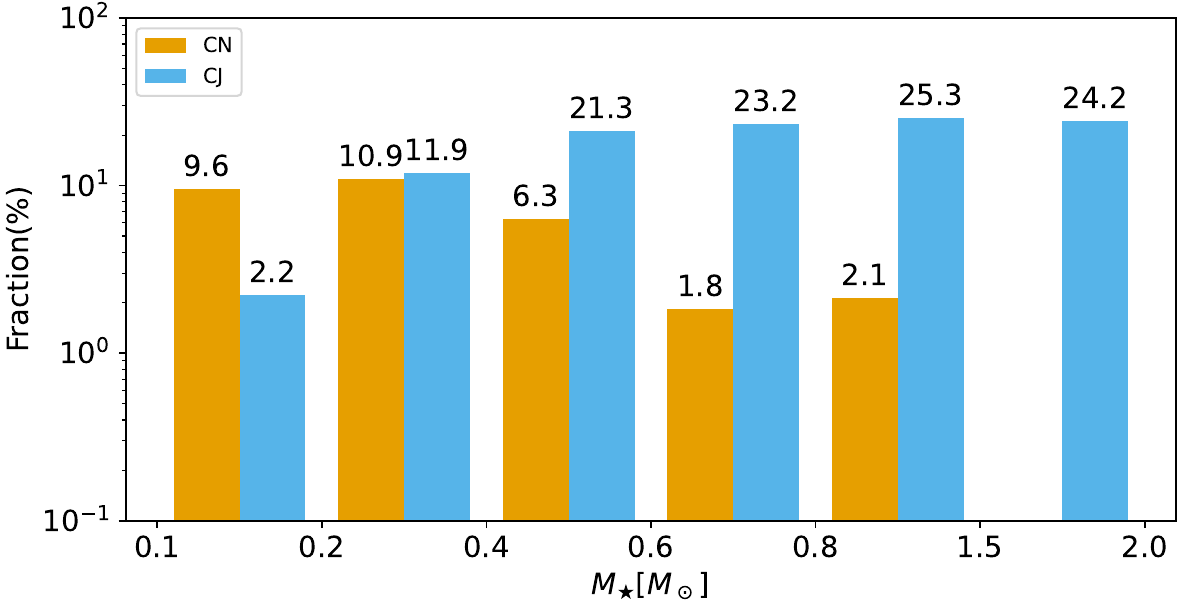}
    \caption{Planet population for isolated stars (left panels) and stars in the cluster (right panels). The initial condition is identical in the two scenarios, while proto-planet disks around stars in the cluster are more massively photo-evaporated during their evolution. In the upper panels, the position of the scatter represents the planetary orbital period and mass at the end of the simulation. The color denotes the host mass. The upper and lower black dashed boxes indicate the boundaries of Jupiter-like and Neptune-like planets, with the associated synthetic fraction labeled. In the lower panels, we separate the occurrence rate of cold Jupiter (blue bar) and cold Neptune (yellow bar) into different host mass ranges. }
    \label{fig:field_vs_cluster_population}
\end{figure*}

The growth of planets around low-mass and high-mass stars exhibit notable differences. On the right axis of the upper panels in \fg{field_vs_cluster_example}, planets initially build their cores efficiently through pebble accretion. Once they reach the pebble isolation mass, gas accretion commences.
The pebble isolation mass is relatively low in the low-mass system (left panel). Consequently, when gas accretion begins, the planet migrates inward because the gas accretion timescale is longer than the Type I migration timescale. While the planet continues to grow, migration slows down due to its ability to gradually open a gap in the disk, thereby reducing migration \citep{KanagawaEtal2018}.
In contrast, the planet in the high-mass system (right panel) has a higher pebble isolation mass, driving faster gas accretion \beq{KH}, and resulting in migration over a much shorter distance. Ultimately, planets in both low-mass and high-mass systems evolve into gas giants within 1 million years.

To examine the impact of external photoevaporation (EPE) on planet formation, we calculated the two stars in the cluster illustrated in \fg{field_vs_cluster_example}. We conduct the simulation of planet formation using identical initial conditions to those in \fg{field_vs_cluster_example} but in a cluster environment. The results of this comparison are presented in \fg{field_vs_cluster_example_massive}. The simulation stops either when the disk is dispersed or at the end of the cluster simulation. 

The insets in the lower panels of \fg{field_vs_cluster_example_massive} illustrate the evolution of the external photoevaporation (EPE) induced by nearby massive stars in the cluster with a time resolution of 10 kyr. The $\dot{M}_\mathrm{EPE}$ evolution curves exhibit discrete feature for two main reasons:
i) Clumps within the molecular cloud shift their relative position to the star, occasionally shielding FUV and EUV luminosity from nearby massive stars.
ii) Close encounters between stars are transient events, contributing to fluctuations in the radiation received. The latter effect matters particularly for low-mass stars, which tend to be further away from the cluster center than the more massive stars due to mass segregation.
 
The disk surrounding the low-mass star is strongly influenced by EPE. In the low-mass system, the $\dot{M}_\mathrm{EPE}$ peaks at $10^{-7}M_\odot\mathrm{yr^{-1}}$ between 0.1 and 0.2 Myr, causing truncation of both the dust and gas components. Between 0.2 and 0.3 Myr, the $\dot{M}_\mathrm{EPE}$ remains at the low value, enabling viscous evolution of the gas disk and the disk outer edge expands. On the other hand, the disk around the high-mass star is more resilient against the EPE. The time-averaged EPE experienced by the low-mass and high-mass systems is indicated by the blue and red crosses in \fg{cluster_epe}, respectively. The more massive system encounters an average EPE approximately 2 times higher than that experienced by the less massive one.
The gas disk around the $0.88\, M_\odot$ star shrinks less, because the disk is more massive initially and the majority of the disk mass is in the outer regions. 

Compared to the isolated systems, the growth of the planet before the pebble-isolation mass is similar in both cases. The pebble-accretion timescale is short ($\gtrsim10$ kyr), during which the disks undergo minimal evolution. 

The planet experiences gas accretion differently in the cluster compared to isolation. Gas accretion lasts for $\gtrsim100$ kyr. \Fg{field_vs_cluster_example_massive} depicts the planet mass in the cluster at the final snapshot of the cluster simulation. By that time, the planet within the low-mass system {gains little gas} due to the strong EPE, which induces a short lifetime of the protoplanet disk. The planet in the high-mass system has evolved into a gas giant, and continues to accrete gas while the gas is available.

The clustered stellar environment leads to distinct planet outcomes compared to planetary systems born in isolation. Low-mass stars have shorter disk lifetimes in clusters, resulting in inefficient planet growth and migration, and therefore more distant rocky {(gas-poor)} planets. In contrast, High-mass stars have longer disk lifetimes and are more resistant to the environment. {On one hand, high-mass stars are born with more massive disks. On the other hand, high-mass stars have higher stellar accretion rates, more materials are transported to the outer disk through viscous processes. These two aspects help the disk radius to shrink less. The resulting planet population in the cluster is similar to the one born in isolation. }

\begin{figure*}
    \centering
    \includegraphics[width=\columnwidth]{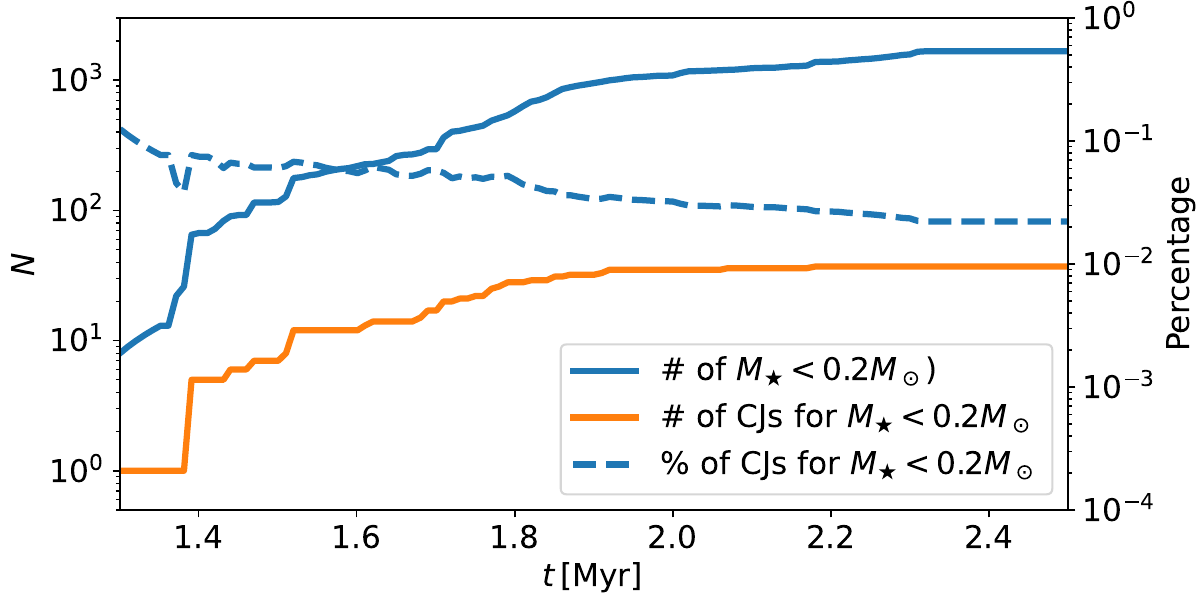}
    \includegraphics[width=\columnwidth]{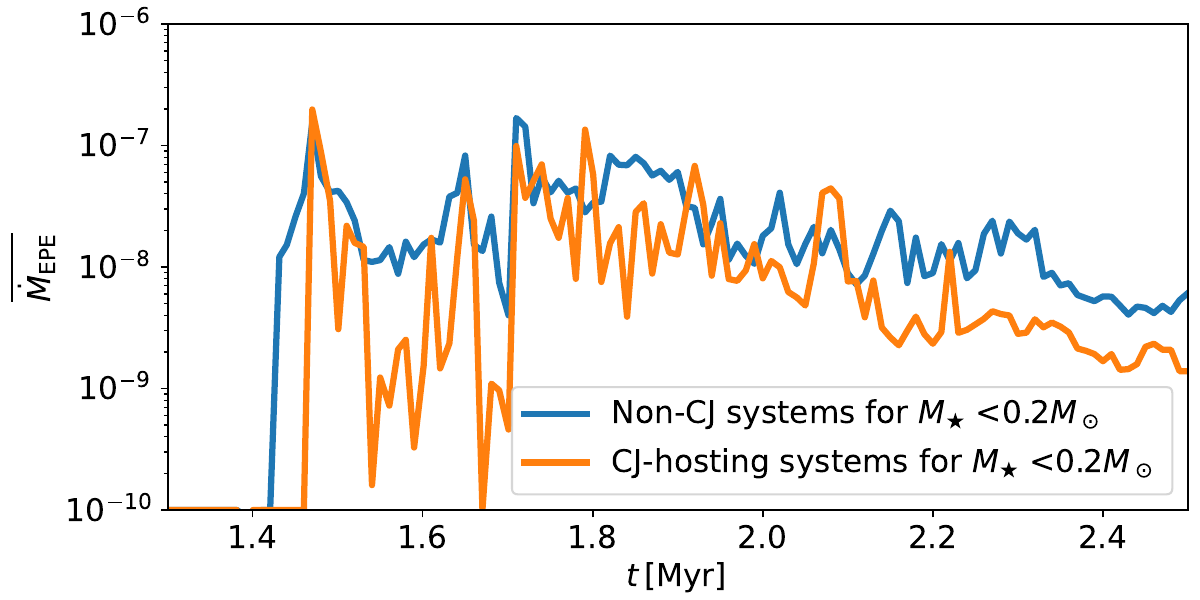}
    \includegraphics[width=\columnwidth]{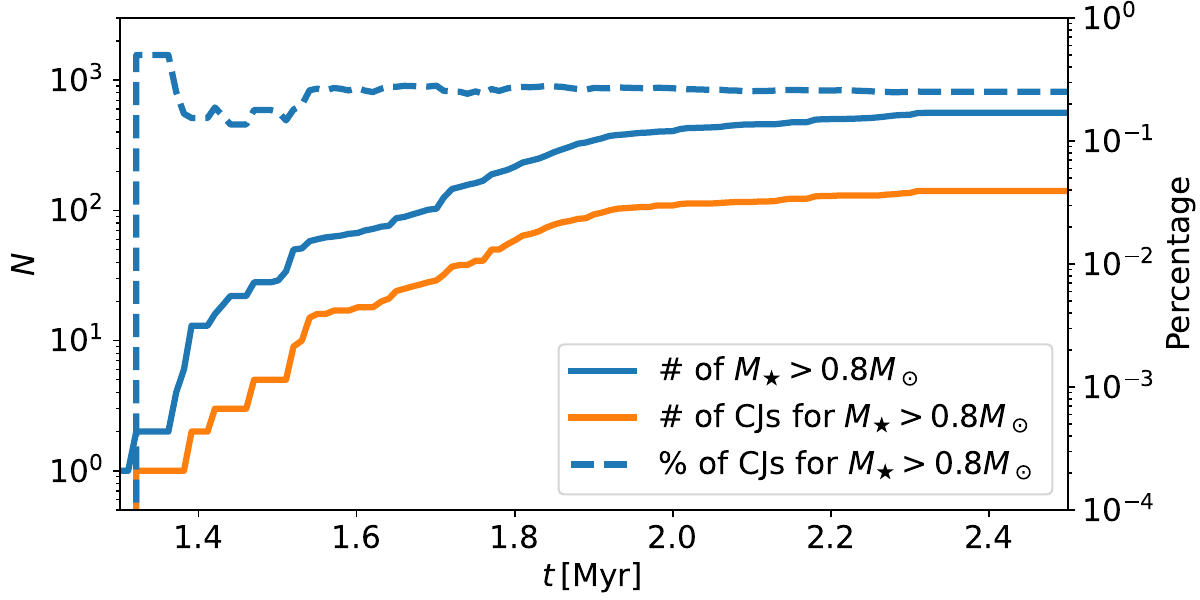}
    \includegraphics[width=\columnwidth]{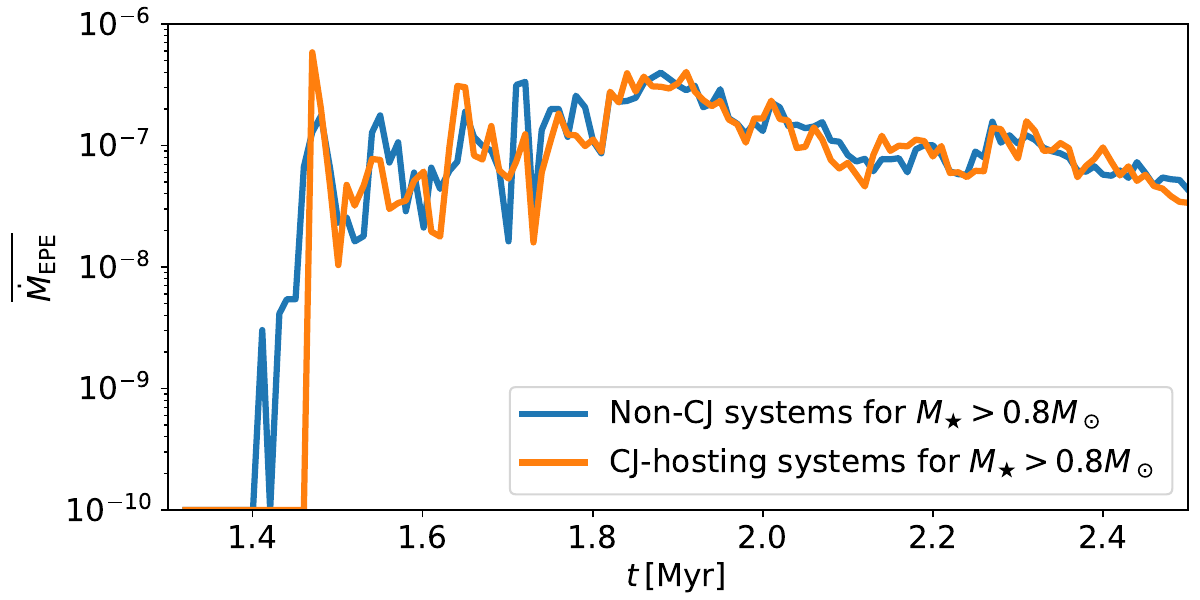}
    \caption{Time evolution of the planet population in the cluster. The top panels illustrate the evolutionary trends of CJ-hosting and non-CJ low-mass systems ($ <0.2M_\odot$), while the lower panels focus on high-mass systems ($>0.8M_\odot$). The total number of low-mass ($ <0.2M_\odot$) and high-mass ($>0.8M_\odot$) systems is depicted with blue solid lines in the left panels. Additionally, the left panels display the number of CJ-hosting systems (orange solid line) and the fraction of CJ-hosting systems (blue dashed line). In the right panels, we compare the system-averaged EPE experienced by non-CJ systems (blue line) and CJ-hosting systems (orange line) just before their CJs are formed. }
    \label{fig:cj-noncj}
\end{figure*}

A recent observation with JWST aligns with our simulations. \cite{BerneEtal2024i} deduced the outward mass flux from the photodissociation regions of the d203-506 disk (a proplyd in the Orion Nebula), situated around an M-star with a mass <$0.3M_\odot$. The estimated mass loss rate is $\dot{M}=1.4\times10^{-7}-4.6\times10^{-6}M_\odot$, indicating a disk lifetime shorter than 0.13 Myr. The short disk lifetime limits the timeframe for gas accretion. The short lifetime of the observed disk is consistent with our simulation of the $0.15M_\odot$ system (left panel of \fg{field_vs_cluster_example_massive}).

\subsection{Planet population: isolated vs clustered}
\label{sec:population}
In the cluster, the EPE varies from star to star and from time to time. A population study is therefore necessary to delineate the statistical behavior of planets formed within the cluster. This section elucidates the distinct planet demographics resulting from isolated stars and those within the clustered environment.

In isolation (upper left panel in \fg{field_vs_cluster_population}), most rocky {(gas-poor)} planets reach the disk's inner edge through Type~I migration. Gas giants, on the other hand, remain in wider orbits because they have much longer migration timescales than terrestrial planets due to the gap-opening process. In addition, most of the rocky planets accrete pebbles and reach pebble isolation mass in the outer orbits, then migrate inward. Given that the pebble isolation mass is higher in the outer regions (see \eq{pebbleiso}), those rocky planets have their mass above the local pebble isolation mass. 

Compared to isolated stars, there are fewer gas giants formed in the star cluster, {especially around low-mass stars} (upper right panel in \fg{field_vs_cluster_population}). In the star-cluster environment, the disk's lifespan is shortened due to intense FUV irradiation. {The lower mass stars have lower mass disks and, therefore have shorter disk lifetime}. Consequently, some {low-mass} systems have their disk lifetime shorter than the gas giant formation timescale. Once the disk disperses, planets are unable to continue gas accretion and fewer gas giants form within the cluster. Furthermore, fewer planets reach pebble isolation mass in the star cluster due to less solid content within the disks than in the isolated environments \citep[cf. ][]{QiaoEtal2023}. 

Planets originating in the cluster environment tend to remain further from the central star compared to those born in isolation. Migration within the disk occurs due to gravitational interactions between the disk and the planet. However, in the cluster environment, many disks evaporate early. Consequently, once the ambient gas disk dissipates, planetary migration ceases as well. As a result, planets in the cluster are inclined to remain in wider orbits.

Given the difference between the demographic discrepancies between planets in isolation and the star cluster, we here define the cold Jupiter (CJ) and cold Neptune (CE) planets as: 
\begin{itemize}
    \item \underline{CJ}: Planets with period time between 1 to 31.6 years (equivalent to 1-10 au if they orbit around a solar mass star) and planet-to-star mass ratio between 0.03\% to 1\% (equivalent to 0.3 to 10 Jovian masses if their hosts' mass is one solar mass).
    \item \underline{CN}: Planets with period time between 1 to 31.6 years (equivalent to 1-10 au if they orbit around a solar mass star) and planet mass between 1 to 50 Earth masses.
\end{itemize}
The motivation of such a definition is elaborated in \se{CJandCN}. In the planet demographic plot (\fg{field_vs_cluster_population} upper panels), we denote CJ and CN planets with upper and lower boxes, respectively, with their occurrence rates labeled. Such a definition of planet populations is advantageous because of their distinct occurrence within and out of the cluster environment. Overall, CJ planets are rare, while CN planets are more abundant in the cluster than in the isolated environment. However, the occurrence rates vary notably across different host masses.

We study the stellar dependency of the planet population by segmenting the stellar population into different mass categories and calculating the corresponding occurrence rates of CJ and CN. The outcomes are illustrated in the lower panels of \fg{field_vs_cluster_population}.
We note a more obvious decline in the occurrence rate of CJ as the host mass decreases. For stars smaller than 0.2 $M_\odot$, the CJ occurrence rate plummets by nearly an order of magnitude. Within the range of 0.2 to 0.4 $M_\odot$ stars, the CJ occurrence rate experiences a twofold decrease. Conversely, CN planets are notably more abundant around low-mass stars within the cluster, with occurrence rates declining as the host mass increases. These CN-harboring systems resemble those depicted in the left panel of \fg{field_vs_cluster_example_massive}. If these systems were to form in an isolated environment, they have more opportunities to accrete gas and evolve into gas giants.

Almost no CNs are formed around isolated stars. Planets there grow and migrate more efficiently than those in the cluster due to longer disk lifetime (especially for M stars). A more detailed comparison between planets in the cluster and isolation is put in \se{detail_comparision}. 

{We finally remark that the simulated planet population exhibits features consistent with observed exoplanet populations. Our simulation produces hot, warm, and cold Jupiters, as well as hot, warm, and cold super-Earths (sub-Neptunes). However, direct comparison with observations is not appropriate for two reasons. Theoretically, our disk model is simplified and does not account for factors such as viscous heating and disk substructures. Observationally, the cold planet population, especially super-Earths, is poorly constrained. Our goal is to study the effect of star clusters on planet formation, and we leave detailed comparisons for future work.}

\subsection{Co-evolution between CJ formation and the cluster}
The young star-forming cluster is a dynamic environment, with its properties constantly changing. The intensity of FUV irradiation experienced by different stars varies strongly across different stages of cluster evolution. So in this section, we analyze planet populations at different times in the cluster evolution.

In \fg{cj-noncj}, we focus on two groups of planetary systems: those with stellar masses $ < 0.2M_\odot$ and those $>0.8M_\odot$. The total number of these systems at different moments in time is represented by the blue solid lines. The shapes of the blue lines in the upper and lower left panels exhibit similar patterns, reflecting the consistent star formation dictated by the same initial mass function (IMF) within the cluster. Overall, star formation decelerates around 1.9 Myr and ceases around 2.3 Myr.


We compare the EPE experienced by CJ-hosting and non-CJ systems around both low and high-mass stars in the right panels of Figure \ref{fig:cj-noncj}. Low-mass systems experience on average lower FUV flux than high-mass systems in the simulation. This is due to the effect of mass segregation. Low-mass stars are more prone to scattering toward larger distances from the cluster center, while their more massive counterparts tend to hardly migrate. Being farther from both the cluster center and the FUV sources, low-mass systems receive lower FUV flux.

Furthermore, for low-mass systems, CJ-hosting systems exhibit lower EPE than non-CJ systems in nearly all the simulation time. In contrast, minimal differences are observed between CJ-hosting and non-CJ systems around high-mass stars. These findings directly highlight the resilience of high-mass systems to EPE irradiation compared to their low-mass counterparts. 

\begin{figure}
    \centering
    \includegraphics[width=\columnwidth]{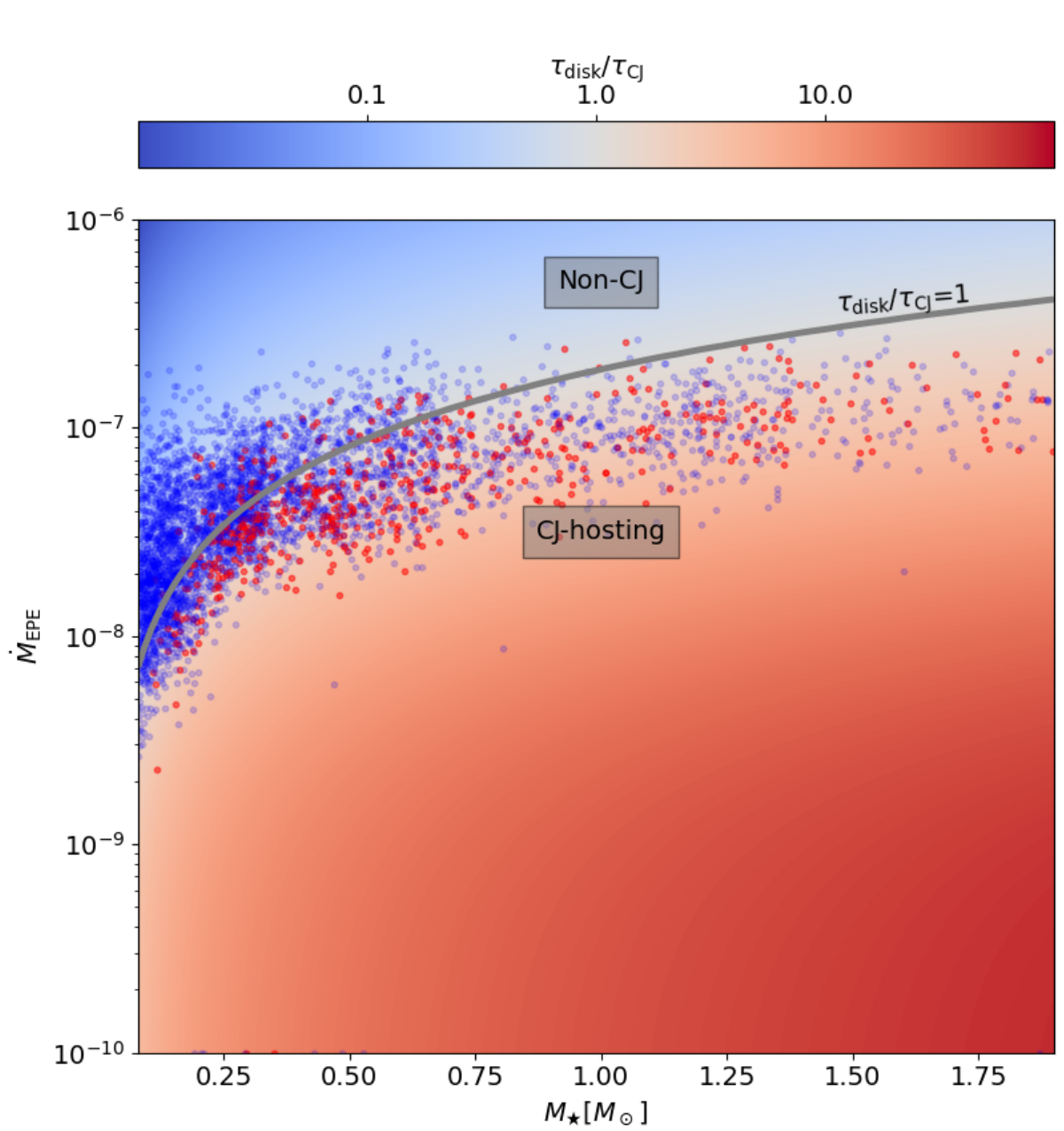}
    \caption{Disk lifetime over CJ growth time with varying host mass $M_\star$ and $\dot{M}_\mathrm{EPE}$. The color represents the values of $\tau_\mathrm{disk}/\tau_\mathrm{CJ}$. The contour line shows where the disk lifetime equals the CJ growth time. CJ-hosting systems are expected to lie below this line, while non-CJ systems lie above it. Our simulation results (\se{default}) are over-plotted, with non-CJ systems shown as blue dots and CJ-hosting systems as red dots. }
    \label{fig:analytical}
\end{figure}

\section{Analytical criteria for CJ formation in the cluster}
\label{sec:theory}
Gas giants can form if the planet's formation timescale {is longer than the disk lifetime}. In this section, we analytically approximate the planet formation timescale at the $P=31.6$ yr orbit. Such orbit is by definition where our outermost CJ is located. The disk lifetime is calculated under different FUV environments. The analytical comparison between the two timescales is then used to validate the simulation outcomes. 

The disk lifetime is determined by both self-accretion and external photo-evaporation processes. In the cluster, the disk mass loss is dominated by the external photo-evaporation rate ($\dot{M}_\mathrm{EPE}$). We can simplify by assuming both rates remain constant over time with $\dot{M}_\mathrm{g}=3\pi \nu \Sigma_\mathrm{g}(t=0)$. The disk's lifetime can then be estimated using the formula:
\begin{equation}
    \tau_\mathrm{disk} = \frac{M_\mathrm{d}}{\dot{M}_\mathrm{g}+\dot{M}_\mathrm{EPE}},
\end{equation}
with the disk mass $M_\mathrm{d}$ proportional to the host mass. It reveals that lower-mass stars have shorter disk lifetimes than higher-mass stars if they are in the same dense cluster environment.

To become a gas giant, a planetesimal first needs to reach pebble isolation mass, then continue to accrete gas. Planet migration complicates this process but is ignored in the approximation. With this simplified model, we can then calculate the timescale for one planet to grow from $0.01$ to $100M_\oplus$.

The pebble accretion rate in the 2D Hill regime \citep{OrmelKlahr2010, LambrechtsJohansen2012} is used for the analytical calculation:
\begin{equation}
\label{eq:Hillpebble}
    \dot{M}_\mathrm{peb} = 2\left(\frac{\mathrm{St}}{0.1}\right)^{2/3}\Omega R_\mathrm{H}^2 \Sigma_\mathrm{p}.
\end{equation}
Here, $\dot{M}_\mathrm{peb}$ relies on the dust Stokes number and surface density. We set $\mathrm{St}=10^{-3}$. Radial drift causes the dust surface density evolves over time. We adopt an approximate analytical form of the dust surface density from \cite{GurrutxagaEtal2024}, which accounts for radial drift and dust growth:
\begin{equation}
    \Sigma_\mathrm{p} = Z_0 T^{-\frac{1}{2-\gamma_\mathrm{t}}\frac{\chi \mathrm{St}}{3\alpha_\mathrm{acc}}} \Sigma_\mathrm{g},
\end{equation}
where $Z_0$ is the initial dust-to-gas ratio, $\gamma_\mathrm{t}$ represents the viscosity index, and $\chi$ corresponds to the negative disk pressure gradient (approximately constant). These values align with the simulation setup outlined in \se{initial}.  We integrate \eq{Hillpebble} and calculate the time $\tau_\mathrm{core}$ needed for one planet to grow its core from $0.01$ to $M_\mathrm{iso}$. 
Then gas accretion commences. At the early stage of gas accretion, Kelvin-Helmholz contraction \beq{KH} dominates. We therefore calculate the time for gas accretion $\tau_\mathrm{gas}$ from $M_\mathrm{iso}$ to $100M_\oplus$ via integrating \eq{KH}. Finally, the CJ growth timescale is:
\begin{equation}
    \tau_\mathrm{CJ} = \tau_\mathrm{core}+\tau_\mathrm{gas}.
\end{equation}
It now allow us to compare $\tau_\mathrm{CJ}$ to $\tau_\mathrm{disk}$. If $\tau_\mathrm{disk}>\tau_\mathrm{CJ}$, CJs can form. In contrast, if $\tau_\mathrm{disk}$ is shorter, the planet can not grow efficiently into CJs. 

We vary the host mass $M_\star$ and $\dot{M}_\mathrm{EPE}$ and calculate the ratio $\tau_\mathrm{disk}/\tau_\mathrm{CJ}$. The results are shown in \fg{analytical}. Generally, $\tau_\mathrm{disk}/\tau_\mathrm{CJ}$ is smaller in the upper left corner and larger in the lower right corner. Low-mass stars, which have initially low-mass disks, exhibit shorter disk lifetimes under the same external conditions. Moreover, the pebble isolation mass is smaller around low-mass stars because their disks are cooler. Consequently, the growth of an embryo into a gas giant takes longer, as the Kelvin-Helmholtz (KH) contraction timescale \beq{KH} is sensitive to the planet's mass initially. Therefore, low-mass stars {need more time }to form gas giants. Our simulation results align with this pattern. {In the regime of low-mass stars, e.g., between 0.08 and 0.25 $M_\odot$, the power law of $\tau_\mathrm{disk}/\tau_\mathrm{CJ}=1$ line exceeds unity. That is because $\tau_\mathrm{CJ}$ negatively correlated with stellar mass and it decreases by a factor of two. }

{In the star clustering environment, CJ planets can form if the disk lifetime is sufficiently long.} {As shown in \fg{analytical},} most simulated CJ-hosting systems have $\tau_\mathrm{disk} > \tau_\mathrm{CJ}$ and relatively low $\dot{M}_\mathrm{EPE}$ compared to non-CJ systems. In the figure, we include the curve for $\tau_\mathrm{disk}/\tau_\mathrm{CJ}=1$. Above this line, the disk evaporates before the gas giant can form, preventing CJ formation. Below the line, CJ formation is possible if the initial planetesimal forms near $P=31.6$ year.

{As a planet grows to Jupiter mass, it begins to open a deep gap in the gas disk, which restricts the amount of material that can enter its Hill sphere (\eq{hillacc}). In this approximated calculation, we only consider KH contraction.}

Our simulation outcomes are over-plotted in \fg{analytical}, with CJ-hosting systems shown as red dots and non-CJ systems as blue dots. The simulation aligns with the analytical results. CJ-hosting systems stay in the lower part with low EPE. The transition from CJ-hosting to non-CJ systems matches $\tau_\mathrm{disk}/\tau_\mathrm{CJ}=1$. For massive stars ($M_\star \gtrsim 1M_\odot$), disk lifetimes exceed the CJ formation timescale. Not all stars successfully cook the CJs because the planets either migrate to the inner hot region of the disk or remain in wider orbits, depending on the initial position of the planet embryos.

In our particular cluster, planet formation around high-mass stars is resistant to the stellar environment, this may change in a denser cluster. Because a denser cluster makes the planet-forming disks closer to the FUV sources (with $F_\mathrm{EPE}\sim \rho_\mathrm{OB}^{2/3}$). For example, in the cluster Westerlund 1, with its mean density of $\sim 3.4\times 10^4$\, stars/pc$^3$ (about three times larger compared to the low-density environment simulated in this paper), we expect fewer CJs formed. {As the disks are in a more irradiative environment. In \fg{analytical}, these stars would be located further up toward the "Non-CJ" region, than our simulated cluster. }

\section{Discussion}
\label{sec:discussion}
\subsection{Model assessment}
We introduce the planet population synthesis code PACE to study planet formation in complex, perturbed environments such as dense stellar clusters. While planets grow within the protoplanetary disks, these disks evaporate due to external photoevaporation from nearby stars in the cluster.

We update the EPE of all disks every 10 kyr using simulation results from {Wilhelm et al. in prep} \citep[same methodology as][]{WilhelmEtal2023}. Note that the pebble accretion timescale is $\gtrsim10$ kyr. 
{While our findings on the planet population generated in the cluster remain qualitatively robust, increasing the time resolution may enhance the precision of planet formation calculations within individual systems.}

\begin{figure}
    \centering
    \includegraphics[width=\columnwidth]{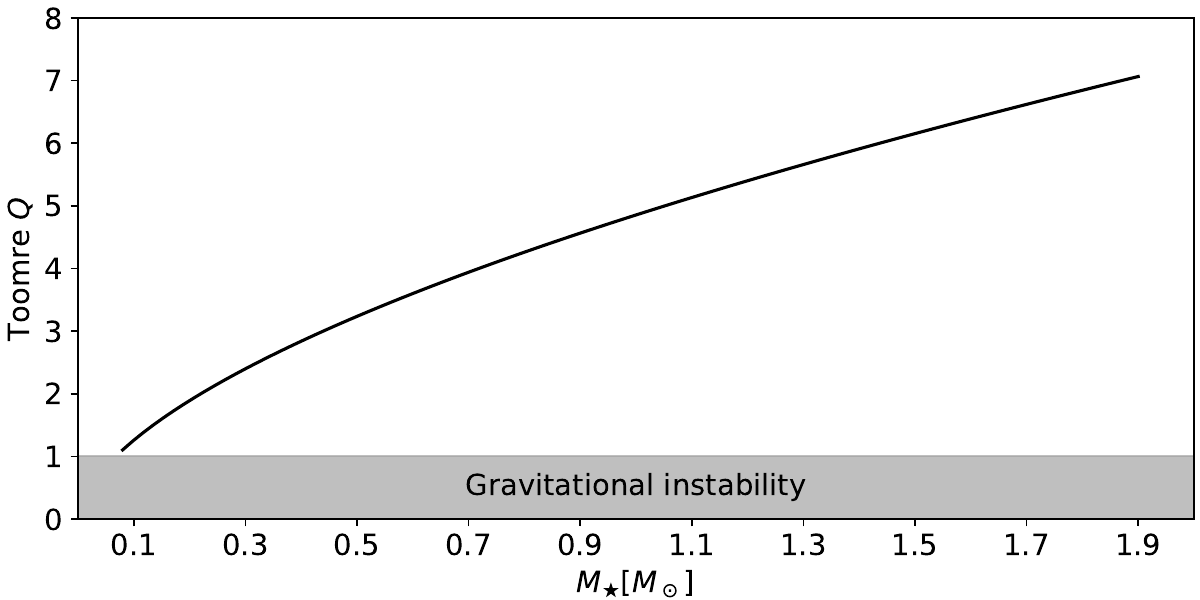}
    \caption{Values of Toomre Q of initial disks at the outer boundary orbit of our cold planet definition ($P=31.6$ yr) for different stellar masses. The lower gray area indicates $Q<1$ where gravitational instability can occur. }
    \label{fig:ToomreQ}
\end{figure}

{The initial disk radius (\eq{rdisk}) and mass (10\% of the host mass) are fixed for a given stellar mass. Although \cite{WilhelmEtal2023} found consistency when comparing the disk radii and mass during the simulation with those of disks observed in dynamic encounters, this simple initialization can introduce some bias to the simulation results. Future studies should use more observationally motivated initial conditions to better explain the observed exoplanet populations. }

The gas disks we initialize are relatively massive, making up 10\% of the host mass. To evaluate whether the gas disk is gravitationally stable, we calculate the Toomre-$Q$ parameter via
\begin{equation}
Q=\frac{C_s\Omega}{\pi G \Sigma_g},
\end{equation}
where $C_s$ is the sound speed. The Toomre-$Q$ values are assessed at $P=31.6$ yr orbits, corresponding to the outermost orbit of our interest for different stars. The results, illustrated in \fg{ToomreQ}, reveal Toomre-$Q$ values exceeding unity across all samples. This indicates that core accretion is the prevailing pathway for forming gas giants in our simulated systems. {We remark that out of $P=31.6$ yr orbit, the disk is gravitationally unstable (especially around low-mass stars) and may form giant planets directly. }

{This work uses a simplified model for disk evolution.} We neglect the dynamical truncation of a protoplanetary disk resulting from stellar flybys. In most cases, external photoevaporation dominates the disk mass loss over dynamical truncation \citep{WilhelmEtal2023}. However, dynamical truncation \citep[e.g.][and references therein]{PortegiesZwart2016i} becomes more important in denser clusters with fewer massive stars (i.e., fewer FUV sources). For instance, less massive stars are formed due to the strong magnetic fields in the nuclear molecular clouds of NGC 1097 \citep{TabatabaeiEtal2018}. The clouds likely fragment, giving rise to numerous low-mass stars. Dynamical truncation then needs to be considered. {Additionally, our model overlooks disk orientation, which may result in different evolution pathways for disks even with exactly the same FUV and EUV environment.}

Nearby massive stars can elevate the outer disk temperature beyond our assumed stellar radiation-dominated disk temperature \citep{Haworth2021}. This temperature increase results in a puffier disk with a larger aspect ratio, which raises the pebble isolation mass \eq{pebbleiso} and reduces the efficiency of pebble accretion. Consequently, the runaway gas accretion phase of the proto-planets may be prolonged.

The formation timescale of cold Jupiters is a critical factor in this study. Recent research suggests that gas accretion begins earlier during the core growth phase \citep{GurrutxagaEtal2024}, potentially shortening the planet's formation timescale. On the other hand, atmospheric recycling could introduce complexity, and therewith prolonging the gas accretion process \citep[e.g.,][]{OrmelEtal2015, MoldenhauerEtal2021}. However, the effectiveness of this process in delaying gas accretion in distant orbits remains uncertain  \citep{WangEtal2023, BaileyZhu2023}. Furthermore, the atmosphere's opacity is also important in the gas accretion phase but uncertain. 

The rapid formation of planetesimals is one key assumption in our model. We assume that when the most massive planetesimal reaches moon mass, the surrounding disk retains the same mass as our initial disk conditions. The optimal initial planetesimal mass is still under debate \citep{SchaeferEtal2017, AbodEtal2019, RowtherEtal2024i}. However, if a planetesimal attains moon mass later in the process {(intuitively, later for planetesimals in the outer orbits)}, there may not be sufficient accretion material available as pebbles rapidly drift inward \citep{LauEtal2024i}. {As a result, the planet accretion/formation timescale is significantly extended. This makes planet formation more susceptible to external photoevaporation.} If disk substructures exist, they may extend the radial drift of pebbles, which would facilitate planet accretion \citep{Morbidelli2020ii, Chambers2021, JiangOrmel2023, GarateEtal2024}.

\subsection{Cold Jupiter and Neptune in observation}
\label{sec:CJandCN}
Our definition of cold planets includes those located between 1 and 10 au. Pebble accretion is particularly effective in this range, where both planetesimal accretion \citep{JohansenBitsch2019, EmsenhuberEtal2021} and gravitational instability (GI) \citep{ForganEtal2018} are less effective.

Our simulations of planet formation on relatively wide orbits in star clusters reveal intriguing systematics that qualitatively align with observational findings:
\begin{enumerate}
    \item {Low-mass stars are born with low-mass disks, therefore generating fewer (less than factor two) giant planets than around high-mass stars in isolation. However, early truncation of gas supply by external photo-evaporation in the star clustering environment leads to even less (an order of magnitude) cold giants around low-mass stars than high-mass stars \fg{field_vs_cluster_population}.} This aligns with the observation that M-stars exhibit fewer Jupiters compared to FGK stars \citep{GanEtal2023}.
    \item In contrast to planet formation around stars in isolation, our simulated planet population in star clusters exhibits a higher abundance of cold super-Earths and Neptunes. Planets on wide orbits are hard to be detected. Microlensing observations suggest that cold Neptunes are prevalent \citep{SuzukiEtal2016, ZangEtal2023}. Potentially, these planets are formed in the star clustering environment.
    \item \cite{GanEtal2023i} observes that the occurrence rate of cold Jupiters around M-stars shows weak dependence on metallicity, in contrast to closer-in planets \citep{LuEtal2020}. A process that is independent of metallicity and favors wide orbits is needed to explain this statistical trend. External photoevaporation (investigated in this work), stellar close encounter \citep{NduguEtal2022}, and extra heating \citep{NduguEtal2018} on protoplanet disks in the young cluster together emerge as plausible explanations due to their independence from disk metallicity and the existence of planets in outer radii.
\end{enumerate} 

The observational statistics mentioned above rely on a limited number of radial velocity and microlensing detections. Future missions, designed to explore wide-orbit planets, such as PLAnetary Transits and Oscillations of stars \citep[PLATO,][]{RauerEtal2014}, Telescope for Habitable Exoplanets and Interstellar/Intergalactic Astronomy \citep[THEIA,][]{SpergelEtal2009}, and ET2.0 \citep{GeEtal2022}, or the upcoming event {Gaia} Data Release 4, jointly hold the promise of providing a more extensive dataset. These missions enable a more accurate characterization of the occurrence rates of cold Jupiters and Neptunes across diverse stellar populations and help validate our findings.


\section{Conclusions}
\label{sec:conclusion}
We designed and built a planet population synthesis code to investigate the impact of an Orion-like star cluster on the planet formation process. With nearby massive stars as FUV sources, the planet-forming disks are prone to external photoevaporation. We study the resulting planet population and compare it to the population born in isolation. 

Our key result is that less cold Jupiters are formed whereas cold Neptunes are more common around M dwarfs in the dense cluster environment. Neptune-sized planets born in isolation always migrate to the disks' inner edge. Planet formation around high-mass stars is less affected by the star cluster. The main reason is that low-mass stars are born with less massive disks and lose mass more rapidly than those around high-mass stars, due to external photoevaporation in the star cluster environment. Consequently, the planets hardly accrete and migrate. In contrast, high-mass disks are more resilient to the cluster environment and gas giants can form.

    
We also derive the critical FUV environment necessary for forming cold Jupiters, which aligns with our population synthesis simulation results.


\begin{acknowledgements}
     The authors thank the referee, Dr. John Chambers, for his thoughtful comments. S.H. would like to express gratitude to Bertram Bitsch for sharing his personal experience with planet population synthesis works. S.H. also thanks Christoph Mordassini, Takayuki Muto and Thomas Haworth for their valuable discussions.
     This work used the Dutch national e-infrastructure with the support of the SURF Cooperative using grant no. EINF-6701.
\end{acknowledgements}

%
%

\bibliographystyle{aa}
\bibliography{ads} 

\begin{appendix}
\section{Planet population differences in isolated stars and a star cluster}
\label{sec:detail_comparision}
We compare isolated planets with those in a cluster, as shown in \fg{fourpanel}. The left panels display the masses and orbital periods of planets born around isolated stars. The right panels show the same for planets born in a cluster. Each planet in the cluster has a corresponding planet in isolation with the same initial conditions. We calculate the mass and period ratios between each planet in isolation and its cluster counterpart (or vice versa), starting from the same initial embryo. 

\begin{figure*}[!ht]
    \centering
    \FloatBarrier
    \includegraphics[width=\columnwidth]{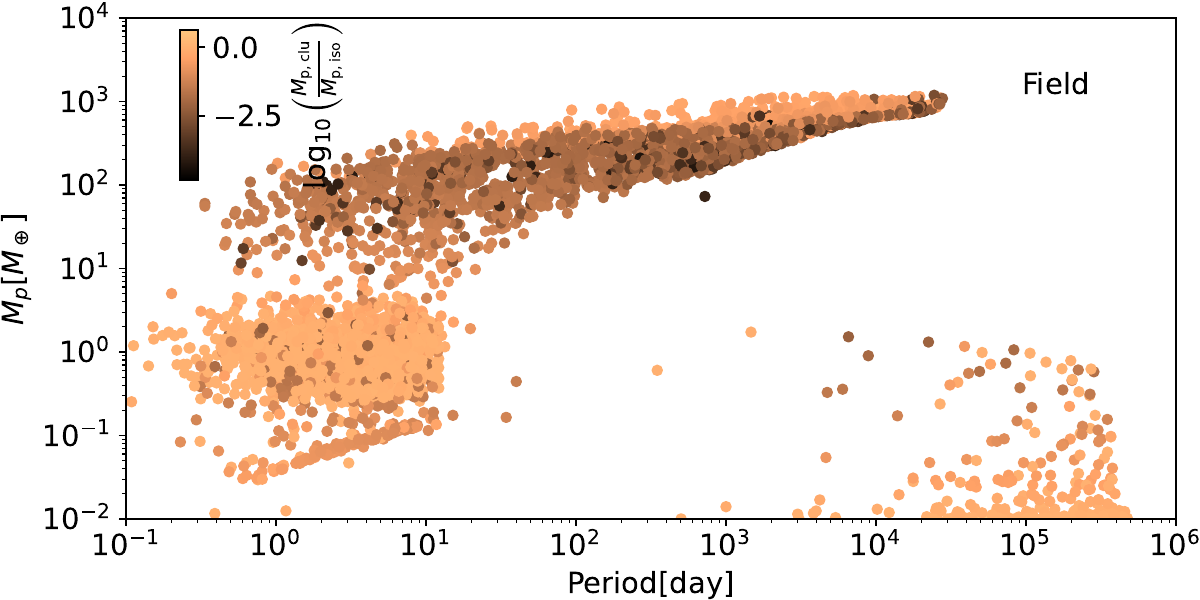}
    \includegraphics[width=\columnwidth]{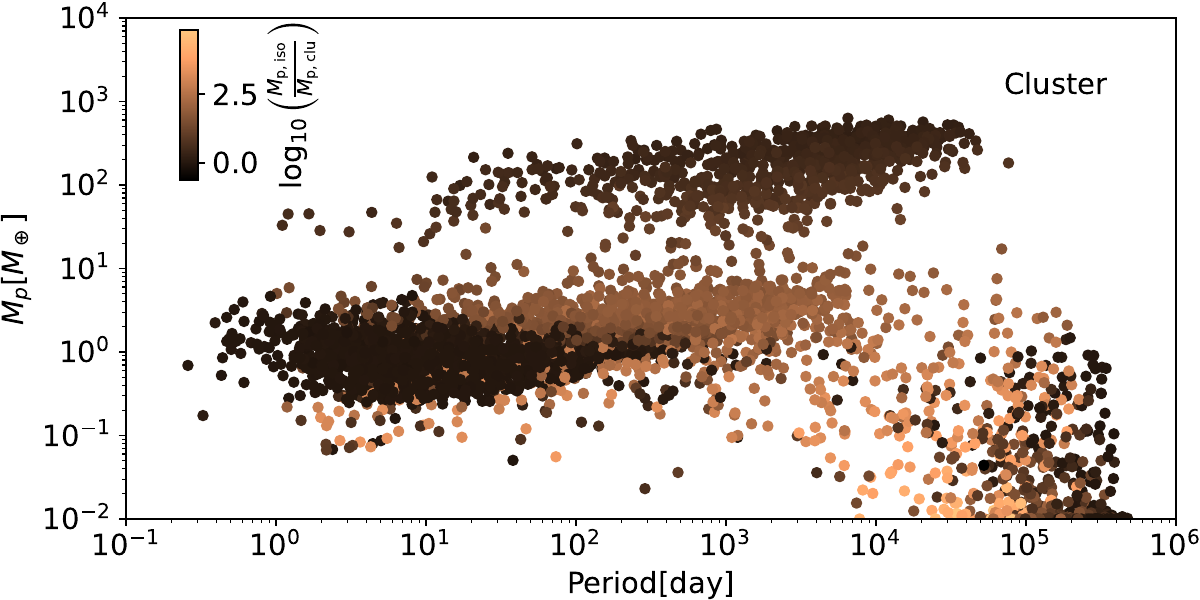}
    \includegraphics[width=\columnwidth]{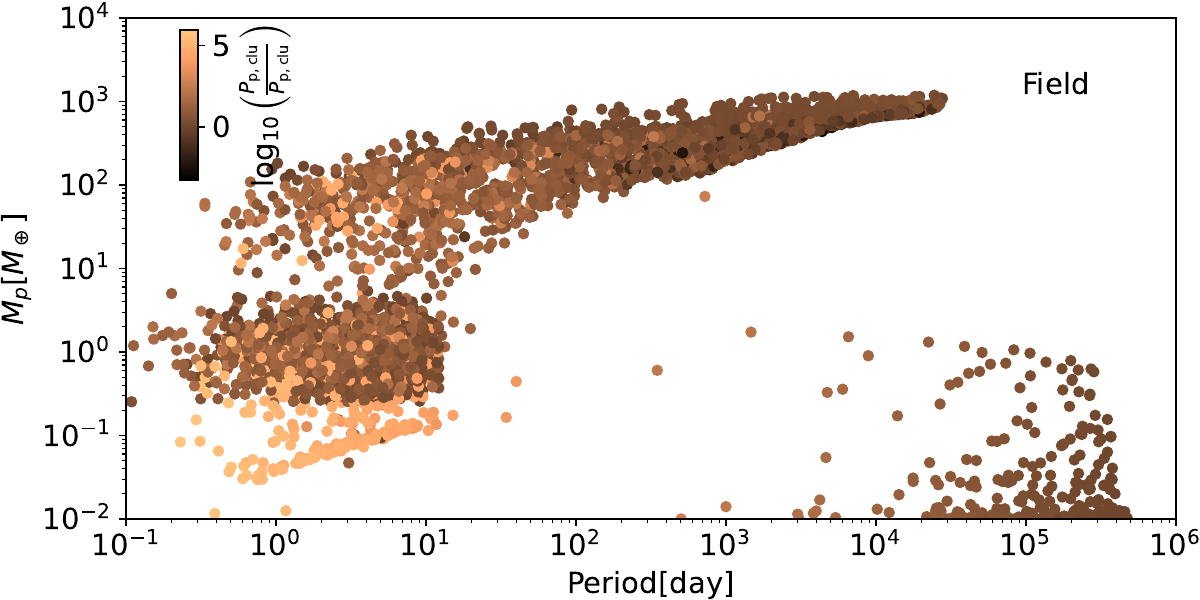}
    \includegraphics[width=\columnwidth]{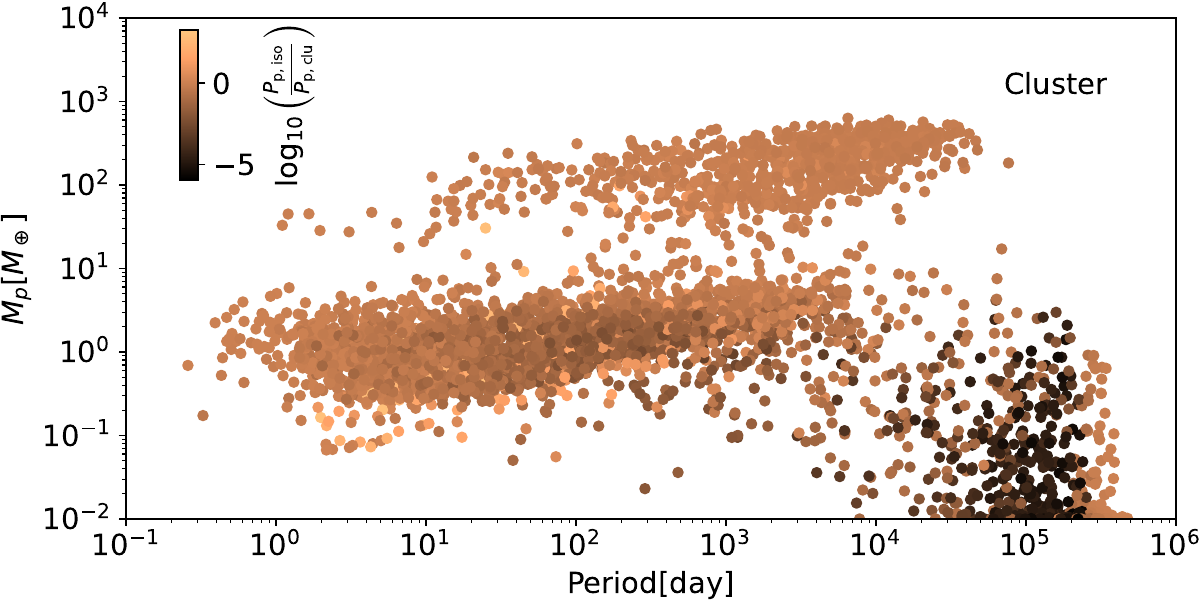}
    \caption{Similar to the upper panels of \fg{field_vs_cluster_population}, but with colors indicating different properties. In the upper left panel, the color shows the logarithmic value of the mass ratio of planets in the cluster to those in isolation. In the upper right panel, it shows the opposite: the mass ratio of planets in isolation to those in the cluster. In the lower panels, the colors represent the logarithmic value of the period ratio of planets. The lower left shows the period ratio of planets in isolation to those in the cluster, while the lower right shows the period ratio of planets in the cluster to those in isolation.  }
    \label{fig:fourpanel}
\end{figure*}

In isolation, planets are always more massive or at least as massive as those formed in the cluster. In both scenarios, rocky {(gas-poor)} planets near the disk's inner edge have similar masses. The most massive gas giants also have similar masses in both scenarios, as they form around massive stars and are more resistant to external environments, as seen in \fg{field_vs_cluster_population}.

In the isolation case, there is a small group of planets around $\sim0.1$ $M_\oplus$ with orbits of 1-10 days. These embryos originate from much wider orbits, between $10^4-10^5$ days. They slowly accrete pebbles until all pebbles drift away, then gradually migrate to the inner edge of the disk. At this location, some pebbles are trapped due to the pressure bump. The embryos then accrete more pebbles and reach the local pebble isolation mass.

As for the embryos starting $\gtrsim10^5$ days, the {pebbles outside the embryos' orbits} are insufficient to grow them up to {$\sim$ Earth mass. Additionally,} their migration timescales are too long. They {therefore} stay close to the initial positions in both isolation and the cluster.

\end{appendix}

\end{document}